\documentclass[a4paper,fleqn,usenatbib]{mnras}
\usepackage{newtxtext,newtxmath}
\usepackage[T1]{fontenc}
\usepackage{ae,aecompl}
\usepackage{graphicx}	
\usepackage{epstopdf}   
\usepackage{amsmath}	
\usepackage{amssymb}	
\usepackage{bm}         
\usepackage{pdflscape}  

\newcommand{\ba}{\begin{eqnarray}}
\newcommand{\ea}{\end{eqnarray}}
\title[Softened potentials of discs]{Potential softening and eccentricity dynamics in razor-thin, nearly-Keplerian discs }

\author[Sefilian \& Rafikov]
{Antranik A. Sefilian$^{1}$
\thanks{E-mail: aas79@cam.ac.uk}
and 
Roman R. Rafikov$^{1,2}$
\\
$^{1}$ Department of Applied Mathematics and Theoretical Physics, CMS, University of Cambridge, Wilberforce Road, Cambridge CB3 0WA, UK
\\
$^{2}$ Institute of Advanced Study, Einstein Drive, Princeton, NJ 08540, USA }

\date{Accepted XXX. Received YYY; in original form ZZZ}
\pubyear{2019}

\begin{document}
\label{firstpage}
\pagerange{\pageref{firstpage}--\pageref{lastpage}}
\maketitle
\begin{abstract}
In many astrophysical problems involving discs (gaseous or particulate) orbiting a dominant central mass, gravitational potential of the disc plays an important dynamical role. Its impact on the motion of external objects, as well as on the dynamics of the disc itself, can usually be studied using secular approximation. This is often done using softened gravity to avoid singularities arising in calculation of the orbit-averaged potential --- disturbing function --- of a razor-thin disc using classical Laplace-Lagrange theory. We explore the performance of several softening formalisms proposed in the literature in reproducing the correct eccentricity dynamics in the disc potential. We identify softening models that, in the limit of zero softening, give results converging to the expected behavior exactly, approximately or not converging at all. We also develop a general framework for computing secular disturbing function given an arbitrary softening prescription for a rather general form of the interaction potential. Our results demonstrate that numerical treatments of the secular disc dynamics, representing the disc as a collection of $N$ gravitationally interacting annuli, are rather demanding: for a given value of the (dimensionless) softening parameter, $\varsigma\ll 1$, accurate representation of eccentricity dynamics requires $N \sim C\varsigma^{-\chi}\gg 1$, with $C\sim O(10)$, $1.5\lesssim \chi\lesssim 2$. In discs with sharp edges a very small value of the softening parameter $\varsigma$ ($\lesssim 10^{-3}$) is required to correctly reproduce eccentricity dynamics near the disc boundaries; this finding is relevant for modelling planetary rings.
\\
\end{abstract}


\section{Introduction} 
\label{sec:intro}

Astrophysical discs orbiting a central mass $M_c$ are ubiquitous in a variety of contexts -- galactic, stellar, and planetary \citep{latter2017}. In many instances, masses of such discs $M_d$ are much less than the central object mass. Despite this fact, gravity of such discs can still play an important dynamical role in the orbital evolution of their constituent particles as well as the dynamics of external objects \citep[e.g.][]{GT79, hep80, war81, koc11, mher12, tey13, mes14, sil15, petrovich18, sef18}. Consequently, characterizing dynamical effects of disc gravity is important.

Whenever $M_d \ll M_c$, particles perturbed by the disc gravity move on nearly-Keplerian orbits which evolve rather slowly. This justifies the use of the so-called \textit{secular approximation} which implies averaging of the fast-evolving dynamical variables over the orbits of particles under consideration \citep{mur99}. The orbit-averaging procedure, also known as Gauss' method, is equivalent to calculating the time-averaged potential due to orbiting point masses by smearing them into massive elliptical "wires" (having shape of their eccentric orbits) with non-uniform linear density proportional to the time spent by an object at a particular phase of its orbit. Such orbit-averaged potential, also known as \textit{secular disturbing function} $R_d$, fully determines the secular dynamics of the system. 

For a test particle with semi-major axis $a_p$, eccentricity $e_p$, and apsidal angle $\varpi_p$ due to a co-planar point mass $\delta m_d$ orbiting with semi-major axis $a$, eccentricity $e_d$, and apsidal angle $\varpi_d$, upon smearing into elliptical rings, the secular disturbing function takes the form \citep{mur99} 
\begin{equation}
\delta R  = \frac{ G \delta m_d a_p}{a^2} \bigg[  \frac{1}{8}b_{3/2}^{(1)} \bigg(\frac{a_p}{a}\bigg) e_p^2  - \frac{1}{4}  b_{3/2}^{(2)} \bigg(\frac{a_p}{a}\bigg) e_p e_d \cos(\varpi_p - \varpi_d) \bigg],
\label{eq:deltaR}
\end{equation}
valid for $a > a_p$ as well as $a < a_p$, as long as particle orbits do not cross. Here $b_s^{(m)}(\alpha)$ is the Laplace coefficient defined by
\begin{equation}
b_{s}^{(m)}(\alpha) = \frac{2}{\pi} \int\limits_{0}^{\pi} \cos(m\theta) \bigg[1+\alpha^2-2\alpha\cos\theta   \bigg]^{-s} d\theta,
\label{eq:bsm}
\end{equation}
which obeys $b_s^{(m)}(\alpha^{-1}) = \alpha^{2s} b_s^{(m)}(\alpha)$. Explicit time independence of $\delta R$ guarantees that the semi-major axes of the secularly interacting objects stay fixed. 

When considering gravitational effects of a razor-thin {\it continuous} disc with smooth distribution of surface density, a straightforward way to compute the secular disturbing function would be to orbit-average the disc potential (obtained by direct integration over its full surface) along the particle orbit. However, this procedure involves a triple integration (two-dimensional integral over the disc surface and orbit averaging) and is numerically challenging.

A more efficient approach lies in representing the disc as a collection of massive, nested, confocal elliptical "wires" (also referred to as "annuli" or "rings" in this work) with fixed semi-major axes \citep[e.g.][]{tou09, bat12}. Due to the additive nature of gravity, the disturbing function due to a disc can be represented as a sum of individual contributions in the form (\ref{eq:deltaR}) produced by all wires, which amounts to integration of $\delta R$ (Eq. \ref{eq:deltaR}) over the radial extent of the disc:
\begin{equation}
R_d = \int_{a_{\rm in}}^{a_{\rm out}} \delta R, 
\label{eq:laplace_Rd}
\end{equation}
where $a_{\rm in}$ and $a_{\rm out}$ are the semi-major axes of the inner and outer disc edges. In this case, provided that $\delta R$ is known as a function of $a$, only a single integration (over the semi-major axes of the rings) is needed, significantly accelerating calculations\footnote{The Laplace coefficients entering in $\delta R$ can be easily evaluated, without relying on integration over $\theta$ in Eq. (\ref{eq:bsm}), by expressing them through elliptic integrals, see Appendix \ref{app:ellipticintegrals}.}. 

Unfortunately, this straightforward procedure is ill-posed from the mathematical point of view. Indeed, it is well known that the Laplace coefficients $b_{3/2}^{(m)}$ featured in Eq. (\ref{eq:deltaR}) diverge as $b_{3/2}^{(m)}(\alpha) \rightarrow (1-\alpha)^{-2}$ when $\alpha\to 1$. This implies that the radial integration in Eq. (\ref{eq:laplace_Rd}) encounters an essential singularity at $a=a_p$. As a result, for a co-planar particle orbiting inside a razor-thin disc, $a_{\rm in} \leq a_p \leq a_{\rm out}$, this direct way of computing $R_d$ does not converge to a finite value. 

This divergence, as well as the pressing need for having an efficient way of computing $R_d$ (via a one-dimensional integration over $a$ only), have motivated the development of alternative analytic approaches for calculating $R_d$. These approaches can be generally grouped into two classes. Calculations of one kind are rooted in the derivation of the potential of an axisymmetric disc with power law surface density profile presented in \citet{hep80}, which does not suffer from the singularity of Laplace-Lagrange secular theory. A number of subsequent studies used this approach \citep{war81} and extended it to the case of eccentric discs, both apsidally aligned \citep{sil15,irina18a} and misaligned (Davydenkova \& Rafikov, in prep.). Higher order (in eccentricity) extensions of this approach have also been developed \citep{sef18}. This framework for treating secular dynamics has been extensively verified using direct orbit integrations under different conditions \citep{sil15, fon16, irina18a}. In this work, we refer to this type of calculation as the unsoftened \textit{Heppenheimer's method}.

Unfortunately, by construction Heppenheimer's method is inapplicable in situations where the disc eccentricity rapidly varies with semi-major axis, potentially resulting in orbit crossings \citep{irina18a}. An alternative approach, which avoids this problem, while at the same time alleviating the aforementioned singularity, is to use \textit{softened gravity} by spatially smoothing the Newtonian point-mass potential in various ways -- both analytically \citep[e.g.][]{tre98, tre01, tou02, hah03, tou12, teygor16} and numerically \citep[e.g.][]{tou09}. In these models, the classical Laplace-Lagrange disturbing function (Eq. \ref{eq:deltaR}) is modified by {\it softening} the interaction potential in some way to circumvent the divergence of $R_d$ as $a \to a_p$. In this method orbit crossing does not lead to problems as long as the softening scale is finite. However, a physical justification for a specific form of softening (absent in the \citet{hep80} approach) often remains unclear, making the introduction of softening rather arbitrary. 

The primary goal of our present work is to assess how well the different calculations relying on potential softening reproduce secular dynamics driven by the gravity of a razor-thin disc. The main metric we use in this exercise is the convergence of the results of such calculations to the true secular evolution (represented by the un-softened Heppenheimer method) in the limit of vanishing softening, when the limit of Newtonian gravity is recovered. Complementary to this, we develop a general framework for computing the well-behaved secular disturbing function for a broad range of softened gravitational potentials.

Our work is organized as follows. We describe the general analytical expressions governing the orbit-averaged potential due to a coplanar disc of arbitrary structure and arbitrary softening prescription in \S \ref{sec:formulation}. Having provided a brief account of the different softened potentials under our probe and the un-softened approach of Heppenheimer in \S \ref{sec:softened_summary} and \S \ref{sec:heppmethod}, respectively, we analyze their performance in reproducing the correct secular dynamics for various disc models in \S \ref{sec:PL}, \S \ref{section:nonPLresults} and \S \ref{sec:edgeeffects}. We discuss and briefly summarize our results in \S \ref{sec:discussion}  and \S \ref{sec:summary} respectively. Technical details of our calculations can be found in Appendices.

\section{Disturbing function due to a disk} 
\label{sec:formulation}

Prior to providing the details of different softening prescriptions examined in this work in \S \ref{sec:softened_summary}, we briefly summarize some of their common features. The ultimate goal of all these prescriptions is the calculation of the disturbing function $R_d$ due to gravity of a (generally eccentric)  disc comprised of massive objects (stars, planetesimals, ring particles) or fluid elements (in gaseous discs) moving on Keplerian orbits. 

We consider the disc to be razor-thin and coplanar. Mass distribution of such a disc can be uniquely characterized by the mass density per unit semi-major axis $\mu_d(a)$, eccentricity $e_d(a)$, and apsidal angle $\varpi_d(a)$ of the trajectories of its constituent elements, as functions of the semi-major axis $a$. In practice, it is often  convenient to use the surface density at periastron $\Sigma_d(a)$ instead of $\mu_d(a)$; its relation to $\mu_d$ for arbitrary profiles of $e_d$ and $\varpi_d$ has been established in \citet{statler2001}, \citet{irina18a} and Davydenkova \& Rafikov (in prep.). Constancy of semi-major axis in secular theory implies that $\mu_d(a)$ does not change in time. The same statement is true for $\Sigma_d(a)$ to lowest order in $e_d$ since $\mu_d(a)  \approx  2 \pi a \Sigma_d(a) + O(e_d)$ \citep{irina18a}.

Close inspection of the various softening methods for computing secular disc potential (\S \ref{sec:softened_summary}) reveals that all of them arrive at the following general form of the disturbing function for a test particle moving on an orbit with the semi-major axis $a_p$, eccentricity $e_p$, and apsidal angle $\varpi_p$: 
\begin{equation}
R_d = {n_p a_p^2}  \bigg[ \frac{1}{2}\mathsf{A}_d(a_p) \mathbf{e}_p^2 + \bm{\mathsf{B}}_d(a_p) \cdot \mathbf{e}_p \bigg].
\label{eq:Rd_general}
\end{equation}
Here $n_p$ is the test-particle mean motion ($n_p^2 = GM_c/a_p^3$), and we have introduced a two-component eccentricity vector for a test particle $\mathbf{e}_p = e_p (\cos\varpi_p, \sin\varpi_p )$.

The coefficients $\mathsf{A}_d$ and $\bm{\mathsf{B}}_d$ in Eq. (\ref{eq:Rd_general}) are related to the disc mass (or surface density) and eccentricity profiles in the following fashion:
\ba
\mathsf{A}_d(a_p) = \frac{2{G}}{n_p a_p^3 }&\times &\Bigg[  \int\limits_{a_{\rm in}}^{a_p} \mu_d(a)  \phi_{22} \bigg(\frac{a}{a_p}\bigg)  da 
\nonumber \\
 & + & \int\limits_{a_{p}}^{a_{\rm out}} \mu_d(a)\frac{a_p}{a} \phi_{11} \bigg(\frac{a_p}{a}\bigg)  da \Bigg],
\label{eq:An_general}
\ea
\ba
\bm{\mathsf{B}}_d(a_p) = \frac{{G}}{n_p a_p^3 }&\times & \Bigg[ \int\limits_{a_{\rm in}}^{a_p} \mu_d(a) \mathbf{e}_d(a)   \phi_{12} \bigg(\frac{a}{a_p}\bigg) da 
\nonumber\\
 & + & \int\limits_{a_{p}}^{a_{\rm out}} \mu_d(a) \mathbf{e}_d(a)\frac{a_p}{a} \phi_{12} \bigg(\frac{a_p}{a}\bigg)  da\Bigg],
\label{eq:An_Bn_general}
\ea
where $\mathbf{e}_d = e_d(a) (\cos\varpi_d(a), \sin\varpi_d(a))$ is the eccentricity vector for an annular disc element\footnote{We refer the reader to \citet{hep80, sil15, irina18a}  for the expressions of $A_d$ and $B_d$ computed using the un-softened Heppenheimer method for different disc models.}.

Functions $\phi_{ij}(\alpha)$, $i,j=1,2$ entering these expressions fully characterize the \textit{softened} ring-ring secular interaction, see Eq. (\ref{eq:R_softened_phi}). They are unique for each potential softening prescription, with explicit forms for the models that we explore in this work specified in Table \ref{table:table1}. This Table shows that coefficients $\phi_{ij}$ appearing in the literature are linear combinations of {\it softened Laplace coefficients} $\mathcal{B}_s^{(m)}$ defined by 
\begin{equation}
\mathcal{B}_s^{(m)}(\alpha,\epsilon) = \frac{2}{\pi} \int\limits_{0}^{\pi} \cos(m\theta) \bigg[1 + \alpha^2 - 2\alpha \cos\theta + \epsilon^2(\alpha) \bigg]^{-s} d\theta.
\label{eq:Bsm_softened}
\end{equation}
The \textit{softening parameter} $\epsilon(\alpha)$ appearing in this definition remains non-zero as $\alpha \to 1$, thus preventing the divergence of the softened Laplace coefficients $\mathcal{B}_s^{(m)}(\alpha,\epsilon)$ at $\alpha = 1$ (unlike the classical $b_s^{(m)}(\alpha)$).
The explicit form of $\epsilon(\alpha)$ is different for every softening method considered in this work, see \S \ref{sec:softened_summary} and Table \ref{table:table1}. Appendix \ref{app:lap} collates some useful relations for softened Laplace coefficients $\mathcal{B}_s^{(m)}(\alpha,\epsilon)$, as well as their approximate asymptotic behavior and relationships to complete elliptic integrals.

The mathematical structure of $R_d$ given by Eq. (\ref{eq:Rd_general}) is similar to that of the classical Laplace-Lagrange planetary theory \citep{mur99}, see Eq. (\ref{eq:deltaR}). Indeed, let us consider mass distribution of a point mass smeared along an elliptical orbit, $\mu_d (a) \rightarrow m_{\rm pl} \delta(a-a_{\rm pl})$ (where $\delta(z)$ is the Dirac delta-function), and set softening to zero (so that $\mathcal{B}_s^{(m)}(\alpha, \epsilon\to 0) \to b_s^{(m)}(\alpha)$). Then one finds that $R_d$ reduces to the un-softened, orbit-averaged potential $\delta R$ due to a planet with mass $m_{\rm pl}$ and semi-major axis $a_{\rm pl}$, with the {\it unsoftened} coefficients $\phi_{ij}$ in the form \citep{mur99}
\begin{eqnarray}
&& \phi_{11}^{\rm LL}(\alpha) = \phi_{22}^{\rm LL}(\alpha) =    \frac{1}{8} \alpha b_{3/2}^{(1)}(\alpha),  
\label{eq:classical_psi1}
\\
&& \phi_{12}^{\rm LL}(\alpha) =  - \frac{1}{4} \alpha b_{3/2}^{(2)}(\alpha),
\label{eq:classical_psi}
\end{eqnarray}
see Eq. (\ref{eq:deltaR}).

Accordingly, it is intuitive to think of Eqs. (\ref{eq:Rd_general})-(\ref{eq:An_Bn_general}) as the continuous version of classical Laplace-Lagrange planetary theory, modified by the introduction of non-zero softening parameter $\epsilon$ to avoid the mathematical divergence of the classical disturbing function as $a \to a_p$.

We emphasize that the functional forms of $\phi_{ij}$ are not simple replacements of $b_s^{(m)}$ appearing in the unsoftened definition (\ref{eq:classical_psi1}) - (\ref{eq:classical_psi}) by $\mathcal{B}_s^{(m)}$. This can be seen in Table \ref{table:table1} where we summarize some of the expressions for $\phi_{ij}(\alpha)$ proposed in the literature and analyzed in this paper (see \S \ref{sec:softened_summary}). Nevertheless, examination of these expressions shows that when $\epsilon^2(\alpha) \to 0$, the coefficients $\phi_{ij}(\alpha)$ do reduce to their unsoftened versions $\phi_{ij}^{\rm LL}(\alpha)$ given by Eqs. (\ref{eq:classical_psi1}) - (\ref{eq:classical_psi}).

In Appendix \ref{app:plummer} we show that the form of the disturbing function given by Eqs. (\ref{eq:Rd_general})-(\ref{eq:An_Bn_general}) is generic for a wide class of softening models (and not just the ones covered in \S \ref{sec:softened_summary}), for which the interaction potential between the two masses $m_1$ and $m_2$ ($m_i \ll M_c$) located at ${\bf r}_1$ and ${\bf r}_2$, correspondingly, relative to the central mass, has a form\footnote{Note that the inter-particle force resulting from such potential does not, in general, obey Newton's third law (as long as $\mathcal{F}(r_1,r_2)\neq$ const).}
\begin{equation}
\Phi_i(\mathbf{r}_1,\mathbf{r}_2) = -{G} m_j \big[ (\mathbf{r}_1 - \mathbf{r}_2)^2 + \mathcal{F}(r_1,r_2)\big]^{-1/2},
\label{eq:FS2}
\end{equation}
with $i,j=1,2$ and $j\neq i$. Here $\mathcal{F}(r_1,r_2)$ represents an arbitrary softening function introduced to cushion the singularity which arises otherwise at null inter-particle separations. Note that in general this potential may depend not only on the relative distance between the two masses $\mathbf{r}_1 - \mathbf{r}_2$, but also on their distances to the dominant central mass $r_1$, $r_2$. 

Explicit demonstration of the connection between the potential (\ref{eq:FS2}) and $R_d$ given by Eq. (\ref{eq:Rd_general}) represents a stand-alone result of this work. In particular, our calculations in Appendix \ref{app:plummer}, which can be skipped at first reading, show that the softening parameter $\epsilon$ featured in the definition (\ref{eq:Bsm_softened}) is related to $\mathcal{F}$ via $\epsilon^2=[{\max(a_1,a_2)}]^{-2}\mathcal{F}(a_1,a_2)$, where $a_{1,2}$ are the semi-major axes of the interacting particles (see Eq. \ref{eq:epsAPP}). The most general expressions of $\phi_{ij}$ entering the arbitrarily softened ring-ring disturbing function, 
\begin{equation}
R_i =  \frac{{G} m_j}{a_>}  \bigg[ \phi_{11}(\alpha) ~ e_1^2  + \phi_{22}(\alpha)~ e_2^2 + \phi_{12}(\alpha) ~ e_1 e_2 \cos(\varpi_1-\varpi_2) \bigg],
\label{eq:R_softened_phi}
\end{equation}
(here  $i = {1,2}$ and $j \neq i$) is given by Eqs. (\ref{eq:Aaverage})-(\ref{eq:Caverage}) in terms of $\mathcal{B}_s^{(m)}(\alpha,\mathcal{F})$. In the above expression, we have defined $a_> = {\rm max}(a_1,a_2)$ and $a_< = {\rm min}(a_1,a_2)$ such that \footnote{Here we clarify that the definitions of $\phi_{11}(\alpha)$ and $\phi_{22}(\alpha)$, even when different (see Table \ref{table:table1} and Appendix \ref{app:plummer}), are swapped upon interchanging $a_1$ with $a_2$ but keeping, by construction, $\alpha = a_</a_> < 1$ -- see Eqs. (\ref{eq:Aaverage}), (\ref{eq:Baverage}) for details.} $\alpha = a_</a_>$.

Note that in equations (\ref{eq:An_general}) and (\ref{eq:An_Bn_general}) we split integration over $a$ in two parts: over the part of the disc interior to $a_p$, and exterior to it. We do this because for some softening functions $\mathcal{F}$ the coefficients $\phi_{ij}(\alpha)$ do not obey certain symmetry properties when $a/a_p$ is replaced with $a_p/a$, see Eq. (\ref{eq:Bsm_inversion}). Moreover, in general $\phi_{11}$ and $\phi_{22}$ are not necessarily identical as in classical Laplace-Lagrange theory (i.e. Eq. \ref{eq:classical_psi1}); see Table \ref{table:table1} and Appendix \ref{app:plummer} for further details. 

As to the physical meaning of $\mathsf{A}_d$ and $\bm{\mathsf{B}}_d$, we remind the reader that $\mathsf{A}_d$ represents the precession rate of the free eccentricity vector of a test particle in the disc potential, while $\bm{\mathsf{B}}_d$ characterizes the torque exerted on the particle orbit by the non-axisymmetric component of the disc gravity. Corresponding forced eccentricity vector is $\mathbf{e}_{p,f} = - \bm{\mathsf{B}}_d/A_d$. In particular, test-particles initiated on circular orbits  experience eccentricity oscillations of maximum amplitude $e_{p}^{m}= 2 \left|\mathbf{e}_{p,f}\right|$.

As $\mathsf{A}_d(a_p)$ and $\mathsf{B}_d(a_p)$ uniquely determine $R_d$ for different forms of softening, comparison of their behavior in the limit of $\epsilon\to 0$ with that found in the unsoftened \citet{hep80} approach (validated in \citealt{sil15, fon16, irina18a}) is sufficient to assess the validity of a particular softening model, see \S \ref{sec:PL}.

\subsection{Summary of existing softening models}
\label{sec:softened_summary}

Here we provide a brief description of the four different softening prescriptions that have been previously proposed in the literature. Corresponding expressions for their  softening parameters $\epsilon^2(\alpha)$ and coefficients $\phi_{ij}(\alpha)$ are provided in Table \ref{table:table1}.

\subsubsection{Formalism of \citet{tre98} -- Tr98}
\label{sect:tremaine}

\citet{tre98} suggested an expression for the secular disturbing function due to a continuous disc, which uses modified Laplace coefficients in the form
\begin{equation}
\mathcal{B}_s^{(m),\mathrm{Tr}} = \frac{2}{\pi} \int\limits_0^{\pi} \cos(m\theta) \bigg[1+\alpha^2-2\alpha\cos\theta + \beta_c^2\bigg]^{-s} d\theta.
\label{eq:Bsm_tremaine}
\end{equation}
Here $\beta_c^2$ is the dimensionless softening parameter, treated as a constant, i.e. independent of distance. The physical interpretation of this manoeuvre is that $\beta_c$, inhibiting the formal divergence of $R_d$ as $a \to a_p$, can be viewed as the disc aspect ratio. Within this prescription, it is intuitive to think of the eccentric "wires" that comprise the disc as having a distance-dependent radius $b=\beta_c \max(a_1,a_2)$. 
In \citet{tre98} coefficients $\phi_{ij}(\alpha)$ were expressed as derivatives of $\mathcal{B}_{1/2}^{(m),\mathrm{Tr}}$ with respect to $\alpha$, see equations (26) of \citet{tre98}.
These expressions, along with their versions modified using the recursive relations for Laplace coefficients (see Appendix \ref{recursive}), can be found in Table \ref{table:table1}.

\subsubsection{Formalism of \citet{tou02} -- T02}
\label{sect:touma}

\citet{tou02} derived the orbit-averaged potential of a disc by assuming individual particles comprising the disc to interact via Plummer potential with a fixed length scale $b_c$ \citep{BT}. Smearing particles into gravitating eccentric wires, \citet{tou02} \citep[see also][]{tou12} derived the expressions (equations (6) of \citet{tou02}) for $\phi_{ij}(\alpha)$ in the form of linear combinations of softened Laplace coefficients $\mathcal{B}_s^{(m),\mathrm{T}}$, similar to those of  \citet{tre98}: 
\begin{equation}
\mathcal{B}_s^{(m),\mathrm{T}} = \frac{2}{\pi} \int\limits_0^{\pi} \cos(m\theta) \bigg[1+\alpha^2-2\alpha\cos\theta + \beta^2\bigg]^{-s} d\theta.
\label{eq:Bsm_touma}
\end{equation}
However, in \citet{tou02} the softening parameter $\epsilon^2(\alpha) = \beta^2$ is no longer a constant but depends on the distance such that $\beta =b_c/\max(a_1,a_2)$. Within this formalism, one can think of a disc as comprised of nested annuli with a constant thickness $b_c$.

\subsubsection{Formalism of \citet{hah03} -- H03} 
\label{sec:hah03method}

\citet{hah03} computed the orbit-averaged interaction between two eccentric wires by accounting for their vertical thickness. The vertical extent $h$ of a ring effectively softens its gravitational potential over a dimensionless scale $H \sim h/a$, which was assumed to be constant in that work \citep[see also][]{war89}. \citet{hah03} demonstrated that the resultant  $\phi_{ij}(\alpha)$ are functions of softened Laplace coefficients
\begin{equation}
\mathcal{B}_s^{(m),\mathrm{H}} = \frac{2}{\pi} \int\limits_0^{\pi} \cos(m\theta) \bigg[1+\alpha^2-2\alpha\cos\theta +  H^2(1+\alpha^2)  \bigg]^{-s} d\theta
\label{eq:Bsm_hahn}
\end{equation}
with constant $H\ll 1$. In other words, the softening parameter is given by $\epsilon^2(\alpha) = H^2(1+\alpha^2)$ in that work. The explicit expressions for $\phi_{ij}(\alpha)$ in terms of $\mathcal{B}_s^{(m),\mathrm{H}}$ are given by equations (17) of \citet{hah03}.

\subsubsection{Formalism of \citet{teygor16} -- TO16} 
\label{sec:explain_gordon}

\citet{teygor16} modified the unsoftened expressions (\ref{eq:classical_psi1}), (\ref{eq:classical_psi}) for $\phi_{ij}^{\rm LL}(\alpha)$ by simply replacing the usual Laplace coefficients $b_s^{(m)}$ with softened versions defined such that 
\begin{equation}
\mathcal{B}_s^{(m),\mathrm{TO}} = \frac{2}{\pi} \int\limits_0^{\pi} \cos(m\theta) \bigg[1+\alpha^2-2\alpha\cos\theta +  S^2 \alpha  \bigg]^{-s} d\theta.
\label{eq:lapgordon}
\end{equation}
Thus, their softening parameter is $\epsilon^2(\alpha) = S^2 \alpha$, where $S$ is a dimensionless constant. According to the authors, this substitution approximates the process of vertical averaging over the disc with constant aspect ratio $S$, and alleviates the classical singularity. The corresponding expressions for $\phi_{ij}(\alpha)$ are given by equations (7)-(9) of \citet{teygor16}. 
\\
\\
The aforementioned softening prescriptions have their softening parameters $\epsilon^2(\alpha)$ controlled by different constants ---  $\beta_c, b_c, H,$ and $S$. For this reason, in what follows -- with some abuse of notation -- we will collectively refer to these constants as ``\textit{softening parameters}" and denote them by $\varsigma$.

\subsection{The unsoftened Heppenheimer method} 
\label{sec:heppmethod}

A different approach to computing the disturbing function of a razor-thin disc has been developed by \citet{hep80}  \textit{without} resorting to any form of softened gravity \citep[see also][]{war81}. The essence of this method is in computing the potential by direct integration over the disc surface before expanding the integral limits (which involve instantaneous particle position $r$) in terms of small eccentricity of a test particle\footnote{Note that the order of these procedures is opposite to what is usual in the Laplace-Lagrange treatment \citep[e.g.][]{mur99}. For further details, see e.g. \citet{hep80}.}. This expansion is followed by time-averaging over the orbit of a test particle. 

The outcome of this procedure is a set of expressions, akin to Eq. (\ref{eq:Rd_general})-(\ref{eq:An_Bn_general}), which are convergent throughout the disc, in contrast to the classical Laplace-Lagrange theory. Mathematically, this convergent behavior is due to the fact that the emergent expressions contain Laplace coefficients $b_{1/2}^{(m)}(\alpha)$ -- and not $b_{3/2}^{(m)}$  -- which diverge only weakly (logarithmically) as $\alpha\to 1$: $b_{1/2}^{(m)}(\alpha) \propto \log(1-\alpha)$. As a result, upon integrating these expressions over the radial extent of the disc, one obtains a convergent and finite result for $R_d$. Physically, convergent expression is only natural since the calculation of the disk potential by direct two-dimensional integration over its surface is fully convergent at every point in the disc. The Heppenheimer's method simply allows one to properly capture this property, unlike the standard Laplace-Lagrange procedure (when applied to continuous discs).

In his pioneering calculation, \citet{hep80} applied this method to axisymmetric power-law discs to recover the orbit-averaged disc potential to second order in eccentricities. This calculation has been subsequently extended to more general disc structures \citep{sil15,irina18a} (hereafter, SR15 and DR18 respectively), as well as to higher order in eccentricities \citep{sef18}. This framework has been extensively verified for eccentric discs using direct integrations of test particle orbits in actual disc potentials \citep[e.g. SR15,][DR18]{fon16}, validating this approach.


\section{Comparison: Power-Law Discs}     
\label{sec:PL}


Our goal is to examine the performance of different softening prescriptions outlined in \S \ref{sec:softened_summary} in comparison with the results obtained using the un-softened Heppenheimer method (\S \ref{sec:heppmethod}). 

We start this exercise using a model of apse-aligned (i.e. $d\varpi_d/da = 0$), truncated power-law (hereafter PL) disc as a simple example. We characterize surface density and eccentricity of such a disc by 
\begin{eqnarray}
\Sigma_d(a) = \Sigma_0 \bigg( \frac{a_{0}}{a}  \bigg  )^p,~~~~~~e_d(a) = e_0 \bigg( \frac{a_{0}}{a}  \bigg  )^q
\label{eq:Sigmad_ed_PL}
\end{eqnarray}
for $a_{\rm in} \leq a  \leq a_{\rm out}$, where $\Sigma_0$ and $e_0$ are the pericentric surface density and eccentricity of the disc at some reference semi-major axis $a_0$. 

Plugging this anzatz into Eqs. (\ref{eq:Rd_general}) -- (\ref{eq:An_Bn_general}), the secular disturbing function $R_d$ due to PL discs can be simplified to \citep{sil15}
\begin{equation}
R_d = K \big[ \psi_1 e_p^2 + \psi_2 e_p e_d(a_p) \cos(\varpi_p - \varpi_d) \big],
\label{eq:Rd_PL}
\end{equation}
where $K = \pi G \Sigma_0 a_{0}^p a_p^{1-p}$ and the dimensionless coefficients $\psi_1$ and $\psi_2$ are given by
\begin{eqnarray}
 \psi_1  &=&  2 \int\limits_{\alpha_1}^{1} \alpha^{1-p} \phi_{22}(\alpha) d\alpha 
 +  2 \int\limits_{\alpha_2}^{1} \alpha^{p-2} \phi_{11}(\alpha) d\alpha, 
 \label{eq:psi1PL}
\\
\psi_2  &=&  2 \int\limits_{\alpha_1}^{1} \alpha^{1-p-q} \phi_{12}(\alpha) d\alpha
 + 2 \int\limits_{\alpha_2}^{1} \alpha^{p+q-2} \phi_{12}(\alpha) d\alpha,     
\label{eq:phi1phi2PL}
\end{eqnarray}
with $\alpha_1 = a_{\rm in}/a_p$ and $\alpha_2 = a_p/a_{\rm out}$.

The coefficients $\psi_1$ and $\psi_2$ are functions of the power-law indices ($p$ and $q$), any softening parameter involved (through $\phi_{ij}$), as well as the test-particle semi-major axis $a_p$ (through $\alpha_{1,2}$). They are related to $\mathsf{A}_d$ and $\mathsf{B}_d$ via 
\begin{eqnarray}
\mathsf{A}_d(a_p)= \frac{2K}{n_p a_p^2} \psi_1,~~~~~~
\mathsf{B}_d(a_p)= \frac{K}{n_p a_p^2}  e_d(a_p)\psi_2.
\label{eq:AdBdpsiPL}
\end{eqnarray}

As shown in Appendix \ref{app:convergence}, for certain ranges of power-law indices $p$ and $q$ both $\psi_1$ and $\psi_2$ converge to values depending only on $p$ and $q$ and a softening parameter used, provided that the test-particle orbit is well-separated from the disc boundaries (i.e. in the limit $\alpha_{1,2} \to 0$). For $p$ and $q$ in these ranges (determined in Appendix \ref{app:convergence} for each of the considered softened formalisms, similar to SR15), the coefficients $\psi_1$ and $\psi_2$ are determined by the {\it local} behavior of $\Sigma_d(a)$ and $e_d(a)$ in the vicinity of test-particle semi-major axis. 

Given this, we first focus on infinitely extended ($\alpha_{1,2} \to 0$) PL discs with $p$ and $q$ within these ranges (we defer discussion of secular dynamics near the disc edges to \S \ref{sec:edgeeffects}). Then, $\psi_1$ and $\psi_2$ become independent of $a_p$ (i.e. functions of $p$, $q$, and $\varsigma$ only), making them useful as simple metrics for judging the validity of different models of softening.

\subsection{Behavior with respect to variation of softening} 
\label{subsubsec:softening}

\begin{figure*}
\includegraphics[width=2\columnwidth]{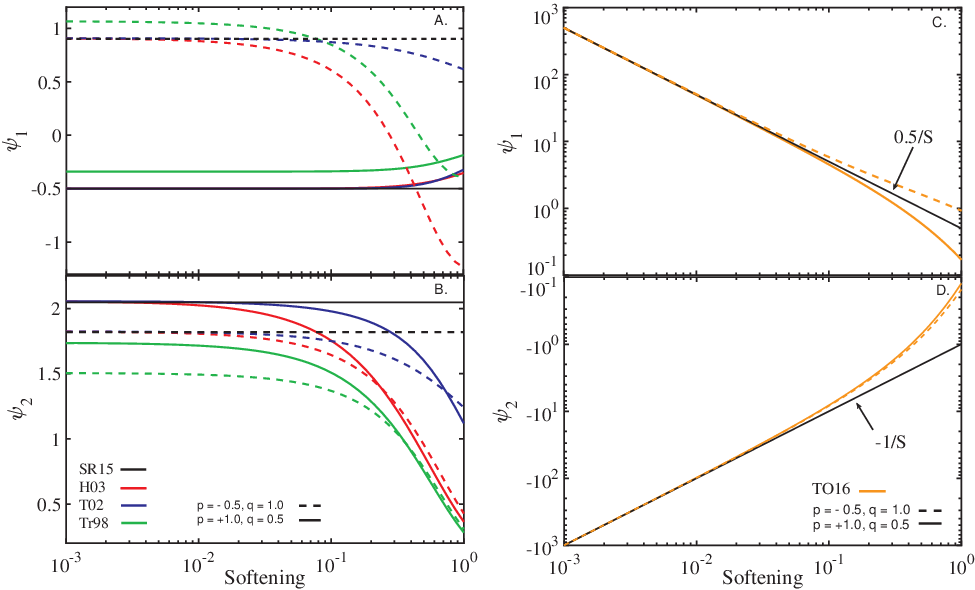}
\caption{ Behavior of the axisymmetric ($\psi_1$, Eq. (\ref{eq:psi1PL}), top panels) and non-axisymmetric ($\psi_2$, Eq. (\ref{eq:phi1phi2PL}), bottom panels) components of the softened gravitational potential due to an infinite power-law disc as a function of softening $\varsigma$. The calculations assume two different disc structures specified by the values of $p$ and $q$ shown by different line types as explained in legend. For clarity, the results obtained by the softened formalisms of \citet{tre98}, \citet{tou02} and \citet{hah03} are collated in the left panels and those obtained by the softening method of \citet{teygor16} are shown in the right panels. The left panels also show the $\psi_1$ and $\psi_2$ obtained by SR15 not assuming any softening (black horizontal lines). See text (\S \ref{subsubsec:softening}) for details. } 
\label{fig:f1}
\end{figure*}

Figure \ref{fig:f1} illustrates the behavior of $\psi_1$ and $\psi_2$ predicted by each of the softening formalisms described in \S \ref{sec:softened_summary} for an infinite PL disc, shown as a function of the corresponding ``softening"\footnote{The softening length $b_c$ present in the formulation of \citet{tou02} is scaled by the test-particle semi-major axis $a_p$ in all the figures where we present results for infinite PL discs. We do this to properly collate the results computed by different softening formalisms in one figure.} $\varsigma$ for two different sets of $p,q$ (indicated in panel B). For reference, black horizontal lines show the values of $\psi_1$ and $\psi_2$ expected from the calculations of SR15 using the un-softened Heppenheimer approach\footnote{Equations (A37) and  (A38) in \citet{sil15} provide analytic expressions for $\psi_1$ and $\psi_2$, respectively, for infinite PL discs.}.

The left panels of Figure \ref{fig:f1} illustrate the behavior of the softening models of \citet{tre98}, \citet{tou02} and \citet{hah03}. They demonstrate that the latter two formalisms predict $\psi_1$ and $\psi_2$ in quantitative agreement with the unsoftened calculations of SR15: results of both \citet{tou02} (blue) and \citet{hah03} (red) converge to the SR15 results as their corresponding softening $\varsigma$ approaches zero; both the amplitude and sign of $\psi_1$ and $\psi_2$ are reproduced. It is also evident that, depending on disc model, $\psi_1$ and $\psi_2$ converge to values given by SR15 at different values of softening. Nevertheless, we generally\footnote{For particles with orbits near sharp disc edges, we find that smaller values of $\varsigma$ is required to recover the expected dynamics, see \S \ref{sec:edgeeffects}.} find that $\varsigma \lesssim  10^{-3}$ guarantees the convergence of $\psi_1$ and $\psi_2$ to within few per cent of the correct values for all $p$ and $q$ as long as $a_{\rm in} \ll a_p \ll a_{\rm out}$ (see Figure \ref{fig:f4}).

The same panels also indicate that $\psi_1(\varsigma)$ and $\psi_2(\varsigma)$ predicted by the softened formalism of \citet{tre98} (green), while converging to finite values as $\varsigma =\beta_c\to 0$, do not reproduce the SR15 results exactly in this limit. Indeed, one can see that even for the smallest adopted value of $\beta_c = 10^{-3}$, the softening prescription of \citet{tre98} yields $\psi_1$ and $\psi_2$ different by tens of per cent from SR15. It is easy to demonstrate that these quantitative differences do not vanish by further decreasing $\beta_c$. For instance, when $p = 1$, the coefficient $\psi_1$ can be evaluated analytically as
\begin{equation}
\psi_1^{\mathrm{Tr98}} = - \frac{1}{2 \sqrt{\beta_c^2+1} } + \frac{\mathbf{E}\bigg(2/\sqrt{\beta_c^2+4}\bigg)}{\pi \sqrt{\beta_c^2+4}} 
= -\frac{1}{2} + \frac{1}{2\pi}  + \mathcal{O}(\beta_c^2) 
\end{equation}
in agreement with Panel A ($\mathbf{E}(k)$ is the complete elliptic integral of a second kind). At the same time, the unsoftened approach of SR15 predicts $\psi_1 = -1/2$ for $p=1$ disc. Moreover, close inspection of Fig. \ref{fig:f1}A,B shows that, in the limit of $\beta_c \to 0$, the $\psi_1$ and $\psi_2$ curves computed using softening model of \citet{tre98} are offset vertically from the unsoftened calculations by  $1/2\pi$ and  $-1/\pi$, respectively, for any $(p,q)$ -- see also Fig. \ref{fig:f4}. We will analyze reasons for this quantitative discrepancy in \S \ref{sec:tremaine_why}.

\begin{figure*}
\includegraphics[width=2\columnwidth]{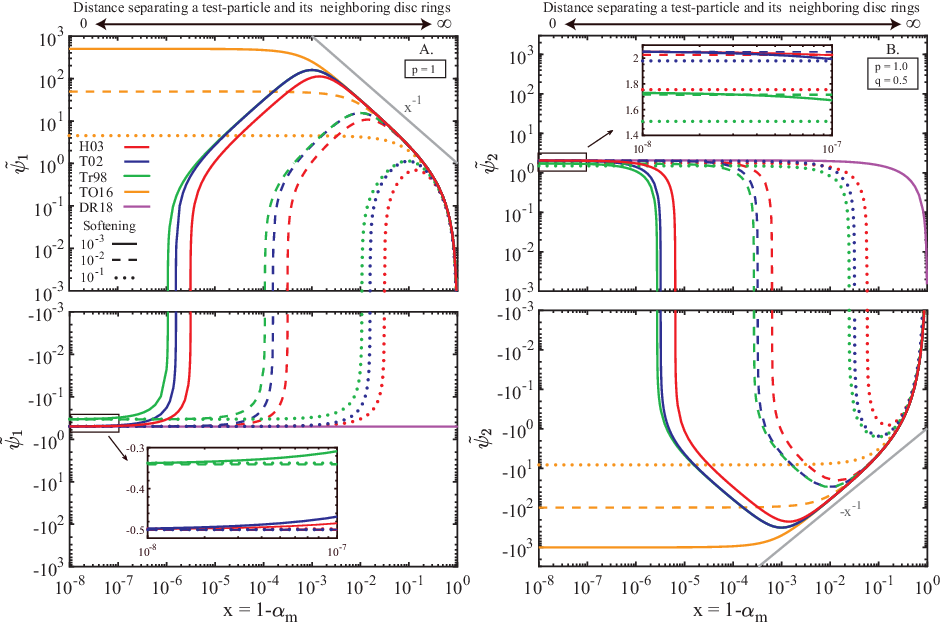}
\caption{Behavior of the cumulative pre-factors $\tilde{\psi}_1(x)$ (panel A) and $\tilde{\psi}_2(x)$ (panel B) of the disturbing function due to a power-law disc ($p = 1$, $q = 0.5$ and $a_{\rm in} \rightarrow 0$, $a_{\rm out} \rightarrow \infty$) with softened gravity, shown as a function of $x$ --- relative separation between a given test-particle orbit and the nearest neighboring disc rings. Formalisms of \citet{hah03}, \citet{tou02}, \citet{tre98} and \citet{teygor16} are shown by different colors as indicated in panel (A), for different values of softening (shown by different line types). The purple lines represent results obtained by the unsoftened expressions of \citet{irina18a} (DR18) based on the Heppenheimer method (see \S \ref{sec:numerical_modeling}). Insets illustrate the behavior as $x \rightarrow 0$ for the three convergent softened formalisms --- see text (\S \ref{ref:spacing_cumulative}) for more details.}
\label{fig:f2}
\end{figure*}

Right panels of Fig. \ref{fig:f1} show the behavior of $\psi_1$ (Panel {C}) and $\psi_2$ (Panel {D}) as a function of ``softening", $\varsigma = S$, resulting from the approach of \citet{teygor16}. There are several features to note here. First, this model predicts $\psi_1>0$ for all values of softening $S$ and disc models (i.e. $p$ and $q$), implying {\it prograde} free precession. This is in contrast with the other softening prescriptions, as well as SR15, which correctly capture {\it retrograde} free precession for $p = 1$ and prograde for $p=-0.5$  (see Panel A). Similarly, $\psi_2$ is always negative, contrary to the expectations (see Panel B). Second, in the limit of $S\to 0$, both $\psi_1$ and $\psi_2$ attain values independent of the disc model, which is clearly inconsistent with the dependence on $(p,q)$ seen in Figure  \ref{fig:f1}A, B. Third, and most importantly, both $\psi_1$ and $\psi_2$ {\it diverge} as the softening $S\to 0$. Indeed, it suffices to employ the asymptotic expansion of the Laplace coefficients $\mathcal{B}_{3/2}^{(m),\mathrm{TO}}$ in the limit of $\alpha \rightarrow 1$ (Eq. \ref{eq:mathcalB_alpha1}) to demonstrate that both $\psi_1$ and $\psi_2$  (Eqs. \ref{eq:psi1PL} - \ref{eq:phi1phi2PL}) behave as 
\begin{equation}
\psi_1^{\mathrm{TO16}}  \approx  \frac{1}{2S} + \mathcal{O}(S),~~~~~~
\psi_2^{\mathrm{TO16}} \approx  - \frac{1}{S} + \mathcal{O}(S)
\label{eq:TO16behav}
\end{equation}
as $S \to 0$ for all values of $p$ and $q$. The behavior shown in Fig. \ref{fig:f1}C, D agrees with these asymptotic expressions. 

\subsection{Details of convergence of different softening prescriptions} 
\label{ref:spacing_cumulative}

Different softening prescriptions explored in this work are designed to modify the behavior of the integrand in equations (\ref{eq:An_general})-(\ref{eq:An_Bn_general}) primarily in the vicinity of the test particle orbit, i.e. as $a\to a_p$ or $\alpha\to 1$. For this reason, it is interesting to look in more detail on how this modification actually allows each softening model to achieve (or not) the expected results. This exercise also illustrates the contribution of different parts of the disc to secular dynamics.

To this goal we compute the values of $\psi_1$ and $\psi_2$ in an infinitely extended PL disc, like in \S \ref{subsubsec:softening}, but now with a narrow clean gap (in semi-major axis) just around the test particle orbit, and explore the effect of varying the width of this gap \citep{war81}. The inner and outer edges of the gap, in which $\Sigma_d(a)$ is set to zero, are at $a_{d,i} =  (1-x) a_p \leq a_p$ and $a_{d,o} = (1-x)^{-1} a_p \geq a_p$, respectively, with a single parameter $x$ controlling the gap width. As $x\to 0$, the width of the gap goes to zero. We compute secular coefficients in such a gapped disc denoted $\tilde{\psi}_{1}(x)$ and $\tilde{\psi}_2(x)$, by appropriately changing the upper integration limits in the definitions (\ref{eq:psi1PL})-(\ref{eq:phi1phi2PL}), i.e. from 1 to $\alpha_m \equiv 1 - x$. This eliminates gravitational effect of the disc annuli with $a_{d,i}(x) < a < a_{d,o}(x)$. 

In Figure \ref{fig:f2} we display the behavior of $\tilde{\psi}_1(x)$ (Panel A) and $\tilde{\psi}_2(x)$ (Panel B) as a function of $x = 1 - \sqrt{a_{d,i}/a_{d,o}}$ for various values of softening $\varsigma$ to highlight the effects of different softening prescriptions. The calculations assume a base PL disc model with $p = 1$ and $q = 0.5$ (recall that $\psi_1$ depends on $p$, while $\psi_2$ depends on $p+q$; Eqs. \ref{eq:psi1PL}, \ref{eq:phi1phi2PL}). There are several notable features in this figure.

First, when the gap is wider than the characteristic softening length $\varsigma a_p$, i.e. $\varsigma \lesssim x \leq 1$, the amplitudes of both $\tilde{\psi}_1(x)$ and $\tilde{\psi}_2(x)$ increase from zero at $x=1$ (infinitely wide gap) to their maximum values reached at $x\sim \varsigma$. In all cases $\psi_1$ is positive, meaning {\it prograde} precession of a test particle orbit in a wide gap, in agreement with the unsoftened results of \citet{war81} and \citet{irina18a} --- secular effect of a collection of distant disc "wires" conforms to expectations of the classical  Laplace-Largange theory (i.e. prograde precession).

In the range  $\varsigma \lesssim x \ll 1$ we find that $\tilde{\psi}_1(x) \sim |\tilde{\psi}_2(x)|\sim x^{-1}$, irrespective of the softening model used; their maximum values are always $\sim \varsigma^{-1}$. This convergent behavior is easy to understand since for $\varsigma \lesssim x$ the role of softening is negligible, $\mathcal{B}_s^{(m)}(\alpha,\varsigma)\approx b_s^{(m)}(\alpha)$,  and all $\phi_{ij}$ effectively reduce to their classical counterparts $\phi_{ij}^{\rm LL}$ given by Eqs. (\ref{eq:classical_psi1}) - (\ref{eq:classical_psi}), which can be easily verified using the expressions listed in Table \ref{table:table1}. The scaling of $\tilde{\psi}_1(x)$ and $|\tilde{\psi}_2(x)|$ with $x$ is simply a result of asymptotic behavior of $b_{3/2}^{(m)}(\alpha) \rightarrow (1-\alpha)^{-2}$ as $\alpha\to 1$, upon radial integration in Eqs. (\ref{eq:psi1PL}) -- (\ref{eq:phi1phi2PL}).

Second, upon reaching their extrema at $x\sim \varsigma$, amplitudes of $\tilde\psi_1(x)$ and $\tilde\psi_2(x)$ computed using softening prescriptions of Tr98, T02 and H03 start decreasing as $x$ decreases. In the range of semi-major axes corresponding to $x\lesssim \varsigma$, softening significantly modifies the behavior of  $\mathcal{B}_s^{(m)}(\alpha,\varsigma)$ away from the divergent behavior of $b_s^{(m)}(\alpha)$. The modification is such that the softened interaction with the disc annuli $\lesssim\varsigma a_p$ away from the test-particle orbit starts to dynamically \textit{counteract} the contribution of the more distant annuli (with $x \approx 1$). As a result of this compensation, $\tilde{\psi}_1$ and $\tilde{\psi}_2$ cross zero and change sign at some $x= C\varsigma^2$, where $C\sim 1$ is a constant\footnote{For  $p=1$, $\tilde{\psi}_1$ becomes analytic for the softened formalisms of both H03 and Tr98 allowing us to quantify the value of $C$. Performing the integral over $d\alpha$ in Eq. (\ref{eq:psi1PL}) - (\ref{eq:phi1phi2PL}), we find $C_{Tr98} = (\pi-1)/2$ and $C_{H03} = \pi$; in agreement with Fig. \ref{fig:f2}. For other values of $p$ and $q$, for which $\psi_1<0$ (c.f. Fig. \ref{fig:f4}), we numerically find that $C$ varies by at most a factor of ten.}. 

At the same time, $\tilde{\psi}_1^{\mathrm{TO16}}$ and $\tilde{\psi}_2^{\mathrm{TO16}}$ calculated according to \citet{teygor16} clearly show different behavior. Instead of decreasing in amplitude as $x\lesssim \varsigma$, they remain essentially constant, having reached their saturated values $\sim \varsigma^{-1}$ at $x\sim \varsigma$. This explains the lack of convergence with $S$ obvious in Figure \ref{fig:f1}C, D, since the values to which $|\tilde{\psi}_1^{\mathrm{TO16}}|$ and $|\tilde{\psi}_2^{\mathrm{TO16}}|$ converge keeps increasing as $\varsigma\to 0$. Moreover, both coefficients also never change sign, always predicting prograde precession ($\tilde{\psi}_1^{\mathrm{TO16}}>0$). The origin of this difference with other smoothing prescriptions will be addressed in \S \ref{sect:gordon}.

Upon further decrease of $x$ below $\varsigma^2$, both $\tilde{\psi}_1$ and $\tilde{\psi}_2$ computed using models of Tr98, T02 and H03 ultimately converge to their corresponding values obtained for a continuous disc (i.e. for $x = 0$, see Fig. \ref{fig:f1}) independent of the assumed value of $\varsigma$. 

We note that the opposite contributions to e.g. $\psi_1$ produced by the distant ($x \gtrsim \varsigma$, positive) and nearby (i.e. with $x \lesssim \varsigma$, negative) disc annuli is not unique to softened gravity. Indeed, both \citet{war81} and \citet{irina18a}, using the un-softened Heppenheimer method, found that a particle orbit fully embedded in a $p=1$ disc has negative precession rate, whereas a particle orbiting fully in the gap precesses in the positive sense (and at high rate if the gap is narrow). As the gap width is reduced, a smooth transition between the two regimes must occur as the test-particle orbit starts crossing the gap edge (i.e. for $x\lesssim e_p$), with the disc annuli crossing the particle orbit giving rise to a negative contribution to $\tilde{\psi}_1$. Eventually, the shrinking of the gap brings $\tilde{\psi}_1$ to a finite negative value (for $p=1$ disc) as $x\to 0$. This sequence is very similar to the behavior we find with softened gravity for $x\lesssim \varsigma$.

\begin{figure}
\includegraphics[width=0.99\columnwidth]{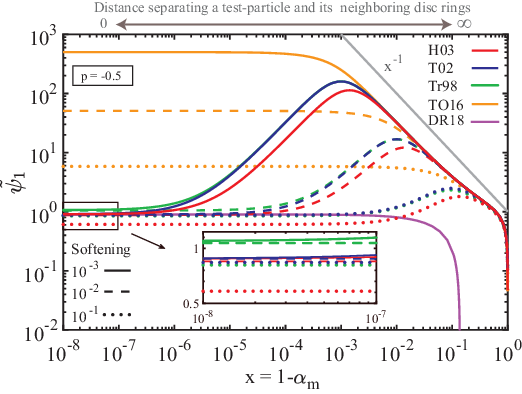} 
   \caption{Same as Figure \ref{fig:f2}, but now for an axisymmetric power-law disc with $p = -0.5$. Note that for this disc model softened $\tilde{\psi}_1(x)$  does not cross zero and converges to a positive value as $x\rightarrow 0$, in agreement with the results in Figure \ref{fig:f1}A.}
   \label{fig:f3}
\end{figure}

In Figure \ref{fig:f3}  we show calculations for $\tilde{\psi}_1(x)$ similar to those in Fig. \ref{fig:f2}A but for a different disc model ---  axisymmetric PL disc with $p = -0.5$. In this case unsoftened calculations (e.g. SR15) predict that disc gravity should drive {\it prograde} precession of a test particle in a smooth disc. One can clearly see that many of the features present in Fig. \ref{fig:f2} are reproduced for this model as well: discrepancy between the TO16 model and others, $\tilde{\psi}_1(x)\sim x^{-1}$ scaling for $\varsigma\lesssim x\ll 1$, decay of $\tilde{\psi}_1(x)$ for $\varsigma^2\lesssim x\lesssim \varsigma$, and ultimate convergence to $\psi_1$ in a disc with no gap. The only obvious difference is the fact that $\tilde{\psi}_1$ does not cross zero\footnote{This is the case for all power-law disc models with $p < 0 $ or $p>3$ for which the expected free precession rate is positive, see Fig. \ref{fig:f4}.} for this disc model with $p = -0.5$.  

To summarize, Figs. \ref{fig:f2}, \ref{fig:f3} indicate that secular dynamics in softened power-law discs is dictated by the delicate balance of the opposing contributions due to nearby (i.e. with $x \lesssim \varsigma$) and distant disc annuli (i.e. with $x \gtrsim \varsigma$), in qualitative agreement with the \textit{unsoftened} results of \citet{war81}. These figures also demonstrate that the softening prescription of TO16 yields inaccurate results due to its inability to capture the dynamical effects of disc annuli adjacent to the test-particle orbit (those with $x \lesssim \varsigma$), see \S \ref{sect:gordon}. We will discuss additional implications of these calculations in \S \ref{sec:numerical_modeling}.

\subsection{Variation of disc model --- $p$ and $q$}
\label{sect:PLvar}

\begin{figure}
\includegraphics[width=1\columnwidth]{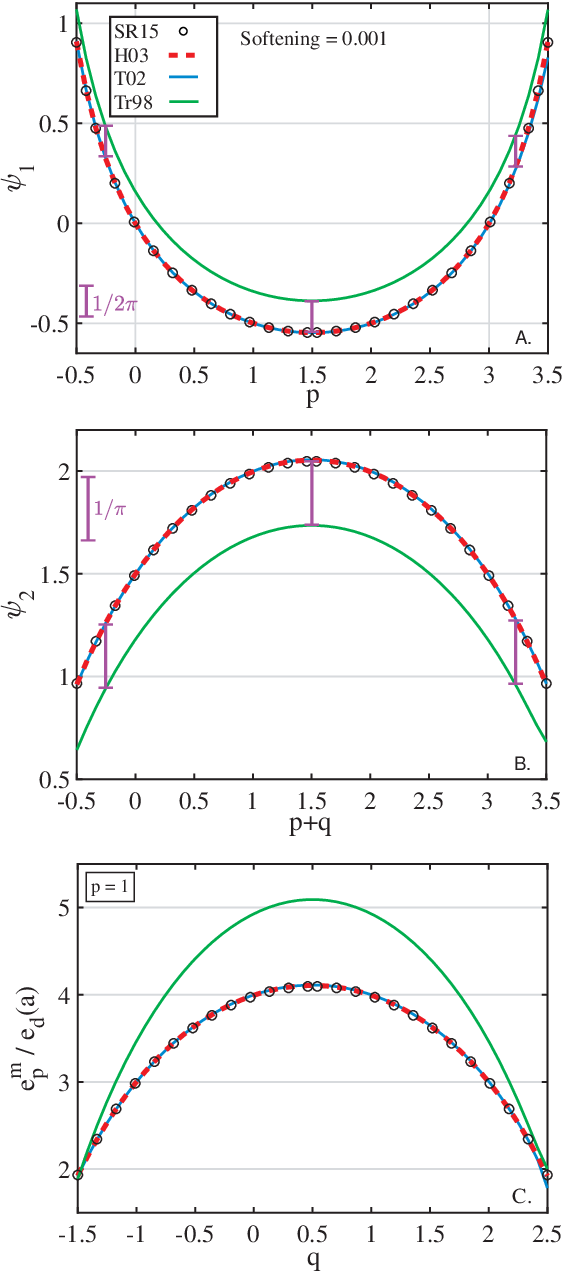}
\caption{
Dependence of the coefficients $\psi_1$ (panel A) and $\psi_2$ (panel B) on the power-law disc model represented by the indices $p$ and $p+q$, respectively. Panel C shows the amplitude $e_p^m$ of eccentricity oscillations (normalized by disc eccentricity $e_d$) induced by disc gravity. Results for softened formalisms of \citet{hah03} (in red), \citet{tou02} (in blue) and \citet{tre98} (in green) are computed using softening $\varsigma = 10^{-3}$. Calculations assume infinitely extended disc (i.e. no edge effects). For reference, open black circles show the profiles of $\psi_1$, $\psi_2$ and $e_p^m$ as computed by SR15: curves for \citet{hah03} and \citet{tou02} fall on top of them, while those for \citet{tre98} show constant offset in terms of both $\psi_{1}$ and $\psi_2$ (illustrated by scale bars in panels A,B) resulting in deviation between $e_p^m$ curves (panel C).
}
\label{fig:f4}
\end{figure}

\begin{figure}
	\includegraphics[width=0.9\columnwidth]{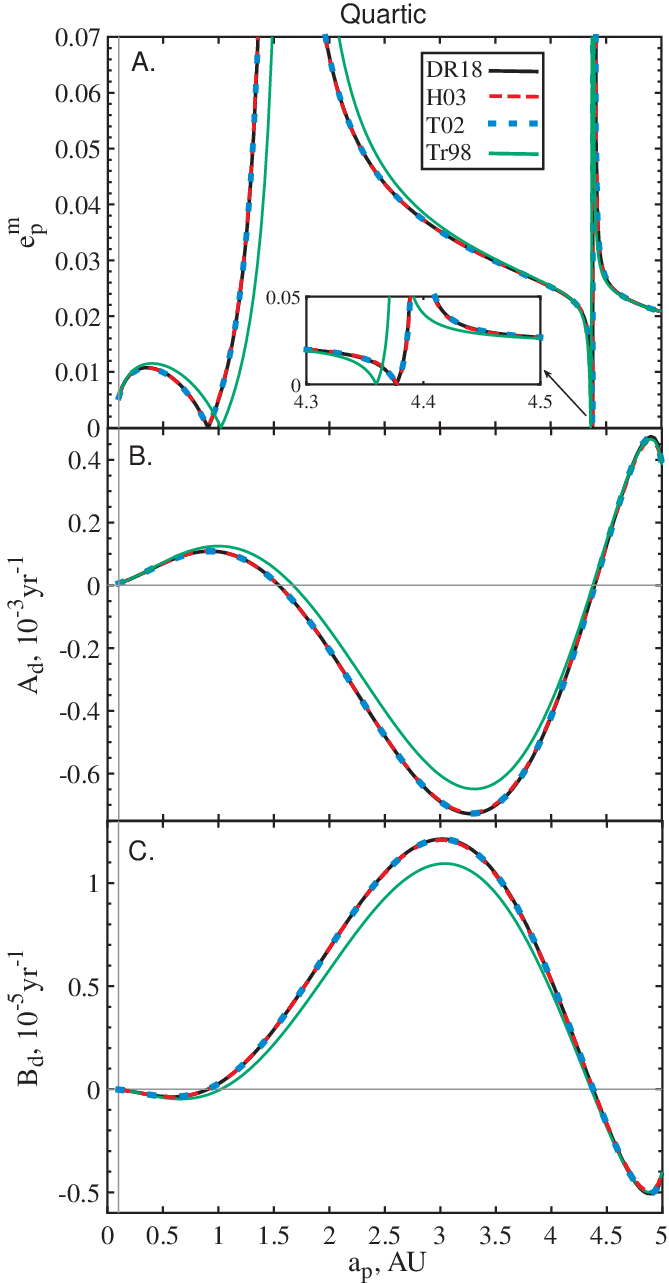}
   \caption{Performance of different softening formalisms (different colors) with softening parameter $\varsigma = 10^{-3}$ in the potential of a Quartic disc, see Eq. (\ref{eq:quartic}), with the eccentricity profile  (\ref{eq:ed_nonPL}). The disc extends from $a_{\rm in} = 0.1$ AU to $a_{\rm out} = 5$ AU. Shown as a function of semi-major axis $a_p$ are the profiles of (A) the amplitude $e_p^m$ of the disc-induced eccentricity oscillations, (B) the rate of disc-driven free precession $A_d$, and (C) the coefficient $B_d$ appearing in the non-axisymmetric part of the disturbing function (\ref{eq:Rd_general}). The black lines represent the expected unsoftened results as computed by \citet{irina18a}. Curves for \citet{hah03} and \citet{tou02} fall on top of the unsoftened results, while the softening method of \citet{tre98} shows only qualitative agreement.}
   \label{fig:f5}
\end{figure}

\begin{figure}
	\includegraphics[width=0.9\columnwidth]{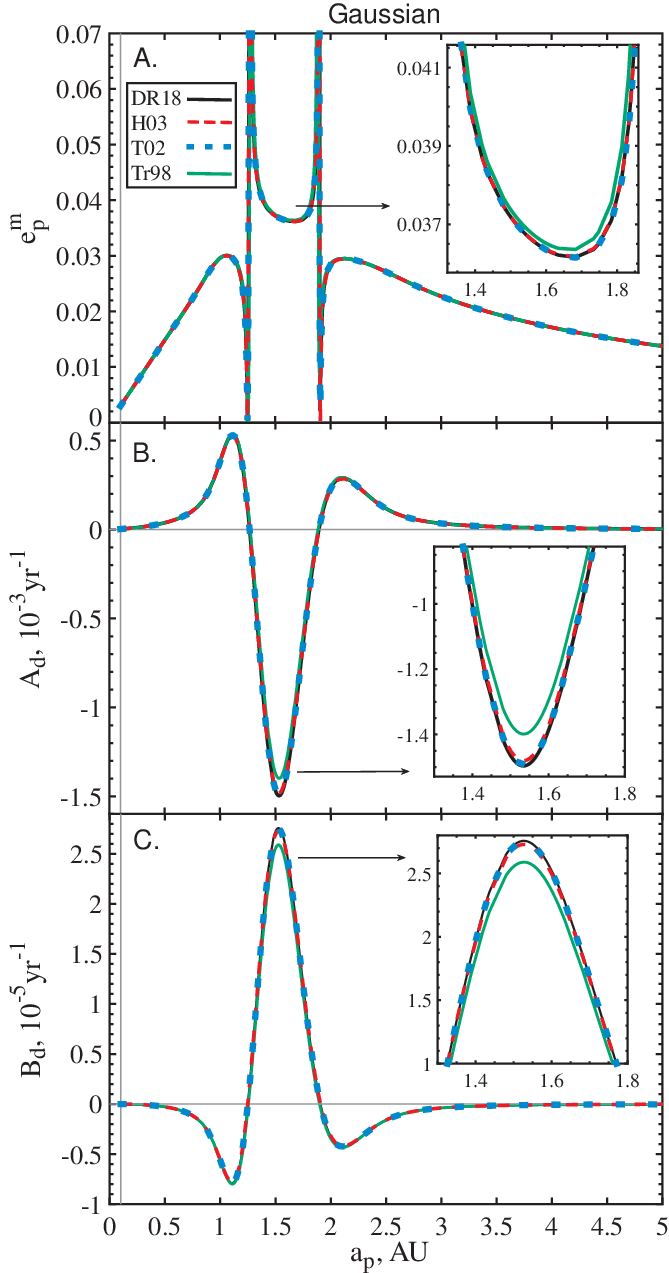}
   \caption{
   Same as Figure \ref{fig:f5}, but now for a Gaussian disc with $\Sigma_d(a)$ and $e_d(a)$  given by Eq. (\ref{eq:GaussianSigma}) and (\ref{eq:ed_nonPL}) respectively.  Note that for this disc model the formalism of \citet{tre98} (green) shows quite good agreement with the unsoftened results, even at the quantitative level. See text (\S \ref{sec:gaussianresults}) for details.
   }
   \label{fig:f6}
\end{figure}

We now examine the dependence of $\psi_1$ and $\psi_2$ on the specifics of the disc model reflected in power-law indices $p$ and $q$. Fig. \ref{fig:f4}A,B illustrates the results based on different softening prescriptions\footnote{We do not present results obtained by the method of \citet{teygor16}.} assuming a softening value of $\varsigma = 10^{-3}$ (for which Fig. \ref{fig:f1}A, B suggests good convergence of $\psi_1$ and $\psi_2$). For reference, black open circles show the expected behavior of $\psi_1$ and $\psi_2$ computed by \citet{sil15} using the un-softened Heppenheimer approach.  

It is clear that the softened formalisms of both \citet{tou02} and \citet{hah03} perfectly reproduce the expected behavior of the pre-factors $\psi_1$ and $\psi_2$ as a function of $p$ and $q$ (i.e. for various PL disc models). On the other hand, the prescription of \citet{tre98} predicts a behavior  of $\psi_1$ and $\psi_2$ only in qualitative agreement with the expected results: the computed values of secular coefficients deviate by tens of per cent from that of SR15. For all values of $p$ and $q$, the formalism of \citet{tre98} yields an additional positive contribution to $\psi_1$ equal to $1/2\pi$ and a negative contribution to $\psi_2$ equal to $-1/\pi$ (these offsets are highlighted in Fig. \ref{fig:f4}A,B by scale bars). Although these differences are not very significant, they lead to (1) predicting a wrong sign for the test-particle free-precession rate for $p \approx 0$ or $p \approx 3$ (for which SR15 yields $\psi_1\approx 0$), and (2) a mismatch of tens of per cent between the  disc-driven forced  eccentricity oscillations, $e_p^m / e_d(a) = | \psi_2/\psi_1|$, and the expectations based on SR15. The latter point is illustrated in Figure \ref{fig:f4}C.

\section{Comparison: non-Power-Law Discs} 
\label{section:nonPLresults}

We now turn our attention to the performance of the different softening prescriptions for more general discs. Namely, we focus on two apse-aligned, non-PL disc models previously studied by \citet{irina18a} based on the unsoftened Heppenheimer method. The dynamics in such non-PL discs, according to DR18, differ from the PL discs in a very important way: the free-precession of test-particles can naturally change from retrograde to prograde (and vice versa) within such discs. Furthermore, an important feature of the models considered below is that $\Sigma_d$ smoothly goes to zero at finite radii in a manner that does not give rise to the edge effects, see DR18 and \S \ref{sec:edgeeffects}.

\subsection{Quartic Disc Model} 
\label{sec:quarticresults}

We start by looking at the secular dynamics in the potential of a {\it Quartic} disc characterized by the surface density 
\begin{equation}
\Sigma_d(a) =  \tilde{\Sigma}_0 \frac{(a_{\rm out}-a)^2 (a_{\rm in} - a)^2    }{ (a_{\rm out} - a_{\rm in})^4   },
\label{eq:quartic}
\end{equation}
and linear eccentricity profile in the form 
\begin{equation}
e_d(a) = \tilde{e}_0 \bigg( 1 + \frac{a_{\rm out}-a }{a_{\rm out}-a_{\rm in}}  \bigg)
\label{eq:ed_nonPL}
\end{equation}
for $a_{\rm in} \leq a  \leq a_{\rm out}$ (with $a_{\rm in} = 0.1$ AU, $a_{\rm out} = 5$ AU), where $\tilde{\Sigma}_0 = 1153$ g cm$^{-2}$ and $\tilde{e}_0 = 0.01$ are normalization constants  (one of the models in DR18). 

Figure \ref{fig:f5} summarizes the salient features of secular dynamics in the potential of such a disc adopting a softening value of $\varsigma = 10^{-3}$. It shows the excellent agreement between the radial profiles of $A_d$, $B_d$ and $e_p^m$ computed using the un-softened calculations of \citet{irina18a} and those computed using softening prescriptions of \citet{tou02} and \citet{hah03}. Similar to the case of PL discs, we find that the softening prescription of \citet{tre98} yields results which agree \textit{qualitatively} with the expected results but differ quantitatively. Deviations of $A_d$ and $B_d$ computed using this model from \citet{irina18a}, in particular, modify the locations at which $A_d$ and $B_d$ become zero. This explains the slight shift in the semi-major axes at which $e_p^m = 2B_d/A_d$ goes through zero or diverges, see Figure \ref{fig:f5}.

The difference between the \citet{tre98} and \citet{tou02} calculations illustrated here could be relevant for understanding the quantitative differences between the studies of \citet{tre01} and \citet{gul12} who analyzed the slow ($m=1$) modes supported by softened Kuzmin discs with softening prescriptions $b \propto r$ and $b = \text{const}$ respectively.


\subsection{Gaussian Rings} 
\label{sec:gaussianresults}


Next we investigate secular dynamics in the potential of another disc model from DR18 --- a Gaussian ring with the surface density profile
\begin{equation}
\Sigma_d(a) = \tilde{\Sigma}_0\exp\bigg\{ \frac{4- [(a/a_c) + (a_c/a)]^2}{w_c}   \bigg\}
\label{eq:GaussianSigma}
\end{equation}
centered around $a_c = 1.5$ AU with width $w_c = 0.18$ and surface density $\tilde{\Sigma}_0 = 100$ g cm$^{-2}$ at $a_c$. The eccentricity profile is still given by Eq. (\ref{eq:ed_nonPL}). 

In Figure \ref{fig:f6} we plot the behavior of the corresponding $A_d$, $B_d$ and $e_p^m$  for the three (convergent) softened formalisms with $\varsigma = 10^{-3}$, together with those of unsoftened Heppenheimer method (DR18, in black). Once again, the results obtained using  the formalisms of \citet{tou02} and \citet{hah03} fall on top of the expectations. However, for this disc model the formalism of \citet{tre98} reproduces the un-softened calculations of \citet{irina18a} quite well: the relative deviations are always less than $10\%$. This improvement will be discussed further in \S \ref{sec:tremaine_why}.

\section{Effects of proximity to the disc edge} 
\label{sec:edgeeffects}

\begin{figure}
	\includegraphics[width=0.99\columnwidth]{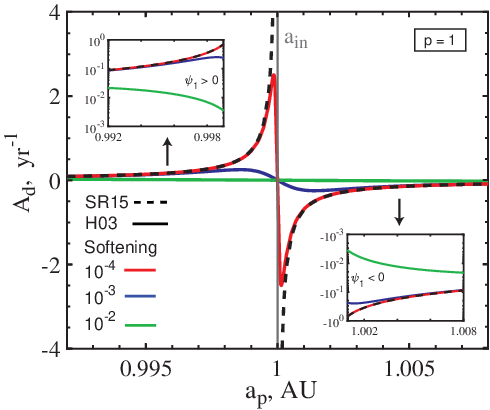}
   \caption{The behavior of the free precession rate $A_d$ near the inner edge $a_{\rm in}=1$ AU of a circular power-law disc with surface density $\Sigma_d(a) = 100$ g cm$^{-2}$ (10 AU/$a)$ (Eq. \ref{eq:Sigmad_ed_PL}). One can see that the expected divergent behavior of $A_d$ near the disc edge is reproduced by the softening prescription of \citet{hah03} in the limit $\varsigma \rightarrow 0$. However, very near the sharp edge of the disc $\varsigma$ has to be very small for quantitative accuracy to be attained. Similar results can be obtained by the softened formalisms of both \citet{tou02} and \citet{tre98}.
   }
   \label{fig:f7}
\end{figure}

So far the disc models that we explored were either infinitely extended (\S \ref{sec:PL}) or had surface density smoothly petering out to zero at finite radii (\S \ref{section:nonPLresults}). This allowed us to not worry about the effects of sharp disc edges --- discontinuous drops of the surface density --- on secular dynamics, which are known to be important \citep{sil15,irina18a}. 

We now relax this assumption and examine the performance of different softening models in the vicinity of a sharp edge of the disc, where surface density drops discontinuously from a finite value to zero at a finite semi-major axis $a=a_{\rm edge}$. To that effect we analyze the behavior of secular coefficient $A_d$ computed using the formalism of \citet{hah03} (we verified that softening prescriptions of \citet{tou02} and \citet{tre98} give very similar results in the limit $\varsigma \rightarrow 0$) for different values of softening (results for $B_d$ are very similar) near the disc edge. Figure \ref{fig:f7} shows the run of $A_d$ near the inner edge $a_{\rm in}$ of the disc for particles both inside ($a_p < a_{\rm in}$) and outside ($a_p > a_{\rm in}$) the disc as predicted by the formalism of \citet{hah03}.  The calculation assumes circular PL disc with $p=1$ and $\Sigma_0 = 100$ g cm$^{-2}$ extending between $a_{\rm in} = 1$ AU to $a_{\rm out} = 10$ AU, where we have set $a_0 = a_{\rm out}$  (Eq. \ref{eq:Sigmad_ed_PL}). 

The unsoftened calculations based on \citet{hep80} invariably predict that the free eccentricity precession rate $A_d$, as well as $B_d$, should diverge as the sharp edge of the disc is approached (e.g. SR15, DR18). \citet{tre01} also found precession rate to diverge near the edge of a Jacobs-Sellwod ring \citep{jac01}. This is indeed the case as shown by the dashed curve computed using SR15. 

The softened calculation using \citet{hah03} does largely reproduce this behavior. However, we find that very close to the ring edge (at $|a-a_{\rm in}|/a_{\rm in}\sim 10^{-3}$) the agreement is achieved only for $\varsigma \leq 10^{-4}$, which is considerably smaller than the values ($\varsigma \sim 10^{-2}$) required to reproduce the dynamics of particles far from the disc edges, $a_{\rm in} \ll a_p \ll a_{\rm out}$, see Fig. \ref{fig:f1}. For $\varsigma = 10^{-2}$ the softened calculation predicts $A_d$ different from the SR15 results near the disc edge by more than an order of magnitude. Thus, accurately capturing secular dynamics near the sharp edges of discs/rings requires using very small values of softening\footnote{On the other hand, this condition is relaxed when the edge is not exactly sharp but rather has a finite width $\Delta r$ over which the disc surface density smoothly peters out to zero; in this case $\varsigma$ only needs to be $\lesssim \Delta r/r$.}. This finding could be problematic, for instance, for numerical modeling of planetary rings, often found to have very sharp edges \citep{Graps95,tis13}.

Note that in Fig. \ref{fig:f7} softened $A_d$ passes through zero exactly at $a_{\rm in}$, showing two sharp peaks of opposite signs just around this radius. Similar behavior was found by  \citet{irina18a} for zero-thickness discs with $\Sigma_d$ dropping sharply but continuously near the edge, demonstrating that variation of the sharpness of the edge is akin to softening gravity. In the case of truly zero-thickness disc and no softening (e.g. SR15) the segment of $A_d$ curve connecting the two peaks turns into a vertical line at $a_{\rm in}$. 

Similar divergent behavior of $A_d$ (and $B_d$) arises also at the outer edge of the disc considered in Fig. \ref{fig:f7} and, in general, at any radius within a disc where $\Sigma_d(a)$ exhibits a discontinuity.
 
Finally, we note that the dynamics of particles orbiting outside the disc (where $\Sigma_d(a) = 0$) is successfully reproduced by the classical Laplace-Lagrange theory without adopting any softening prescription \citep[e.g. see][]{petrovich18}. Indeed, outside the radial extent of the disc semi-major axis overlap (i.e. $a_p = a$) is naturally excluded thus avoiding the classical singularity. Outside the disc the unsoftened calculations based on the Heppenheimer method (e.g. SR15, DR18) reduce to the Laplace-Lagrange theory exactly.

\section{Discussion} 
\label{sec:discussion}

Results of previous sections reveal a diversity of outcomes when different softening models are applied. Two models --- those of \citet{hah03} and \citet{tou02} --- successfully reproduce the un-softened calculations based on the Heppenheimer method in the limit of zero softening. In the same limit, the formalism of \citet{tre98} yields convergent results which are, however, different from the un-softened calculations, typically by tens of per cent. Finally, the softening method of \citet{teygor16} does not lead to convergent results in the limit of vanishing softening parameter. Interestingly, the two successful models \citep{hah03,tou02} have been derived using rather different underlying assumptions (see \S \ref{sect:touma} \& \ref{sec:hah03method}), producing different mathematical expressions for $\phi_{ij}$ (see Table \ref{table:table1}), and yet their results are consistent with the un-softened calculations as $\varsigma\to 0$. 

To understand this variation of outcomes, we developed a general framework for computing secular coefficients $\phi_{ij}$ (thus fully determining the softened secular model via Eqs. (\ref{eq:Rd_general})-(\ref{eq:An_Bn_general})) given an arbitrary softened two-point interaction potential in the form (\ref{eq:FS2}). This procedure involves orbit-averaging the softened potential along the particle trajectories; its details are presented in Appendix \ref{app:plummer}. There is also an alternative approach, sketched in Appendix \ref{app:disc}, which assumes the disc to be a continuous entity from the start. Both of them arrive at the same expressions for $R_d$.

Using these results we show in Appendix \ref{app:correctgordon} that the expressions for $\phi_{ij}$ found by \citet{tou02} and \citet{hah03} can be recovered exactly using this general framework if we set $\mathcal{F}(r_1,r_2) = b_c^2$ and $\mathcal{F}(r_1,r_2) = H^2(r_1^2+r_2^2)$, respectively, in the expression (\ref{eq:FS2}) for the two-point potential. This approach also allows us to address some of the questions raised above, which we do in \S \ref{sec:tremaine_why} \& \S \ref{sect:gordon} below.


\subsection{On the softening prescription of \citet{tre98}} 
\label{sec:tremaine_why}


Results of \S \ref{sec:PL} \& \S \ref{section:nonPLresults} indicate that the softening prescription of \citet{tre98} -- unlike that of \citet{tou02} and \citet{hah03} -- leads to quantitative differences compared to the un-softened calculations. We now demonstrate where these differences come from. 

The form of the softened Laplace coefficient $\mathcal{B}_s^{(m),\mathrm{Tr}}$ defined by Eq. (\ref{eq:Bsm_tremaine}) suggests interaction potential (\ref{eq:FS2}) with $\mathcal{F}(r_1, r_2) = \beta_c^2 {\rm max}(r_1^2,r_2^2)$ for the softening model of \citet{tre98}. In Appendix \ref{app:correctgordon} we show that propagating this form of $\mathcal{F}(r_1, r_2)$ through our general framework results in the following expressions for the coefficients $\phi_{ij}$:
\begin{align}
\phi_{11} = \phi_{22}
&=  \frac{\alpha}{8}  \bigg[ \mathcal{B}_{3/2}^{(1), \mathrm{Tr}} -3 \alpha \beta_c^2 \mathcal{B}_{5/2}^{(0),\mathrm{Tr}}  - \delta(\alpha-1) \beta_c^2 \mathcal{B}_{3/2}^{(0),\mathrm{Tr}}  \bigg], 
\label{eq:new_Tr_axi}
  \\
\phi_{12} &= -\frac{\alpha}{4}  \bigg[  \mathcal{B}_{3/2}^{(2),\mathrm{Tr}}  -3 \alpha \beta_c^2  \mathcal{B}_{5/2}^{(1),\mathrm{Tr}} - \delta(\alpha-1) \beta_c^2 \mathcal{B}_{3/2}^{(1),\mathrm{Tr}}  \bigg].
\label{eq:new_Tr_ecc}
\end{align}
These expressions are different from the entries in the Table \ref{table:table1} for \citet{tre98} in a single but very important way --- presence of terms involving Dirac $\delta$-function. Such terms arise because the form of $\mathcal{F}(r_1, r_2)$ adopted in \citet{tre98} is not sufficiently smooth --- its first derivative is discontinuous at $r_1=r_2$, while the calculation of $\phi_{ij}$ involves second-order derivatives of $\mathcal{F}$, see Eqs. (\ref{eq:T123})-(\ref{eq:T678}), as well as Eq. (\ref{eq:disc_Ad}). Such singular terms do not arise in other types of softening prescriptions examined in our work since they all use infinitely differentiable versions of $\mathcal{F}(r_1, r_2)$. Thus, these terms should not be interpreted as representing some kind of ``self-interaction" within the disc, they merely reflect the mathematical smoothness properties of $\mathcal{F}$ used in \citet{tre98}.

Presence of these terms in Eqs. (\ref{eq:new_Tr_axi})-(\ref{eq:new_Tr_ecc}) introduces corrections to coefficients $A_d$ and $B_d$ (Eqs. \ref{eq:An_general}, \ref{eq:An_Bn_general}) in apse-aligned discs in the form
\begin{eqnarray}
\delta A_d(a_p) &=& -\frac{\pi G}{2 n_p a_p} \beta_c^2 \Sigma_d(a_p) \mathcal{B}_{3/2}^{(0),\mathrm{Tr}}\bigg|_{\alpha=1}, 
\label{eq:AdcorrTr}
\\
\delta B_d(a_p) &=& +\frac{\pi G}{2 n_p a_p} \beta_c^2 \Sigma_d(a_p) e_d(a_p) \mathcal{B}_{3/2}^{(1),\mathrm{Tr}}\bigg|_{\alpha=1}.
\label{eq:BdcorrTr}
\end{eqnarray}
Accounting for these corrections, we confirmed that the correct (un-softened) behavior of the coefficients of $R_d$ can be reproduced for the non-PL discs -- Quartic and Gaussian models, see \S \ref{section:nonPLresults}.
Note that $\delta A_d(a_p)$ and $\delta B_d(a_p)$ are proportional to the local disc surface density $\Sigma_d(a_p)$ and $\mathcal{B}_{3/2}^{(m),\mathrm{Tr}}(\alpha=1) \sim \beta_c^{-2}$, see Eq. (\ref{eq:mathcalB_alpha1}). This likely explains the improved agreement between the calculations of \citet{tre98} and \citet{irina18a} for Gaussian rings (see Fig. \ref{fig:f6}), which feature mass concentration in a narrow range of radii (in contrast to the Quartic model, see Fig. \ref{fig:f5}). 

For PL discs the terms proportional to $\delta$-function in Eqs. (\ref{eq:new_Tr_axi})-(\ref{eq:new_Tr_ecc}) give rise to corresponding modifications of the coefficients $\psi_1$ and $\psi_2$ defined by Eqs. (\ref{eq:psi1PL})-(\ref{eq:phi1phi2PL}):
\begin{eqnarray}
\delta \psi_1 &=& -\frac{1}{4}\beta_c^2 \mathcal{B}_{3/2}^{(0),\mathrm{Tr}}\bigg|_{\alpha=1} = -\frac{1}{2\pi} +O(\beta_c^2),
\\
\delta \psi_2 &=& \frac{1}{2}\beta_c^2 \mathcal{B}_{3/2}^{(1),\mathrm{Tr}}\bigg|_{\alpha=1} =  \frac{1}{\pi} +O(\beta_c^2),
\end{eqnarray}
see Eqs. (\ref{eq:AdBdpsiPL}). These corrections exactly match the offsets seen in Fig. \ref{fig:f4} between the calculations of \citet{tre98} and the un-softened calculations, thus explaining the origin of these uniform shifts. We also confirmed this explanation in Fig. \ref{fig:f8}, where we show the convergence of modified \citet{tre98} coefficients to the correct un-softened values as softening is varied for 2 values of $p$ and $q$.

To summarize, Eqs. (\ref{eq:new_Tr_axi})-(\ref{eq:new_Tr_ecc}) should replace the expressions given by Eq. (26) of \citet{tre98} {\it in applications to continuous discs}. However, when considering the interaction of two individual annuli with different semi-major axes (like in the classical Laplace-Largange theory), one has $\alpha\neq 1$ and terms in Eqs. (\ref{eq:new_Tr_axi})-(\ref{eq:new_Tr_ecc}) containing $\delta$-function naturally vanish, reducing $\psi_1$ and $\psi_2$ back to the expressions quoted in \citet{tre98}.

\subsection{On the softening prescription of \citet{teygor16}}
\label{sect:gordon}

\begin{figure}
	\includegraphics[width=1\columnwidth]{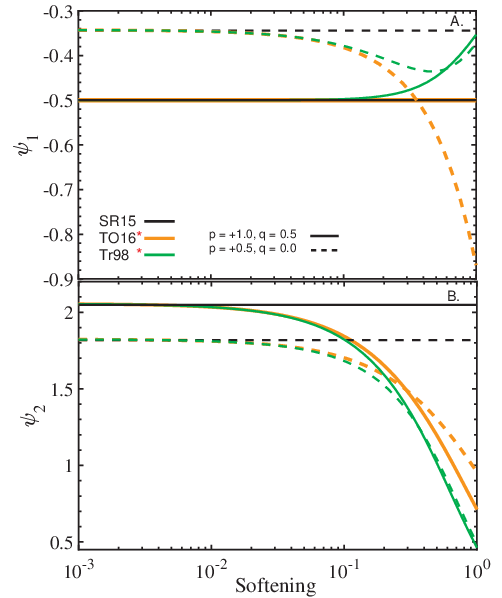}
   \caption{
Similar to Figure \ref{fig:f1}, but now using the expressions for $\phi_{ij}$ given by Eqs. (\ref{eq:new_Tr_axi}-\ref{eq:new_Tr_ecc}) and Eqs. (\ref{eq:new22}-\ref{eq:new2}) obtained by propagating $\mathcal{F}(r_1,r_2) = \varsigma^2 \rm{max}(r_1^2,r_2^2)$  of \citet{tre98} and
$\mathcal{F}(r_1,r_2) = \varsigma^2 r_1 r_2$ of \citet{teygor16}, respectively, through the general framework outlined in Appendix \ref{app:plummer}. Shown as a function of softening $\varsigma$ are $\psi_1$ (panel A) and $\psi_2$ (panel B) for two PL disc models specified by $p$ and $q$ indicated in panel A. Black lines represent the expectations based on \citet{sil15}, to which the new expressions for $\psi_1$ and $\psi_2$ successfully converge as $\varsigma \to 0$.}
\label{fig:f8}
\end{figure}

We now turn our attention to the model of \citet{teygor16} trying to understand its distinct (divergent) behavior. From the expression for $\mathcal{B}_s^{(m),\mathrm{TO}}$ in Eq. (\ref{eq:lapgordon}) one infers that this model features softening parameter in the form $\epsilon^2(\alpha) = S^2 \alpha$. To soften secular interaction \citet{teygor16} directly substituted $b_{3/2}^{(m)}$ in the classical expressions (\ref{eq:classical_psi1}, \ref{eq:classical_psi}) for $\phi_{ij}^{\rm LL}$ with $\mathcal{B}_{3/2}^{(m), \mathrm{TO}}$, see \S \ref{sec:explain_gordon}; this simple swap of Laplace coefficients has not been justified rigorously. 

On the other hand, in Appendix \ref{app:correctgordon} we show that softening parameter in the form $\epsilon^2(\alpha) = \varsigma^2 \alpha$ corresponds to softening function $\mathcal{F}(r_1, r_2) = \varsigma^2 r_1 r_2$ in the two-point potential (\ref{eq:FS2}), see Eq. (\ref{eq:epsAPP}). Propagating such a form of $\mathcal{F}(r_1,r_2)$ through our general framework in Appendix \ref{app:plummer}, we find the following expressions for the coefficients $\phi_{ij}$ with $\varsigma=S$ (Appendix \ref{app:correctgordon}):
\begin{align}
\phi_{11} &= \phi_{22} \nonumber \\
&=  \frac{\alpha}{8}  \bigg[ \mathcal{B}_{3/2}^{(1), \mathrm{TO}} + \frac{1}{2}S^2 \mathcal{B}_{3/2}^{(0),\mathrm{TO}}  
- \frac{3}{4}S^2(2+2\alpha^2+S^2\alpha)\mathcal{B}_{5/2}^{(0),\mathrm{TO}} \bigg], 
\label{eq:new22}
  \\
\phi_{12} &= -\frac{\alpha}{4}  \bigg[  \mathcal{B}_{3/2}^{(2),\mathrm{TO}} + \frac{1}{2}S^2 \mathcal{B}_{3/2}^{(1),\mathrm{TO}}  - \frac{3}{4}S^2(2+2\alpha^2+S^2\alpha)\mathcal{B}_{5/2}^{(1), \mathrm{TO}} \bigg].
\label{eq:new2}
\end{align}

Approach of \citet{teygor16} accounts for only the first terms in Eqs. (\ref{eq:new22}), (\ref{eq:new2}), with coefficients which are $O(S^0)$, see Table \ref{table:table1}. However, as we show below, the correct behavior of $\phi_{ij}$ as $S\to 0$ is guaranteed only when {\it all} the terms present in the above expressions are taken into account.

To demonstrate this, in Figure \ref{fig:f8} we repeat the same convergence study as in \S \ref{subsubsec:softening} but with the modified $\phi_{ij}$ given by Eqs. (\ref{eq:new22}) -- (\ref{eq:new2}). One can see see that the correct implementation of the softening $\epsilon^2(\alpha) = S^2 \alpha$ proposed by TO16 leads to the recovery of the expected test-particle dynamics in infinite PL discs; this is very different from the divergent behavior obvious in Fig. \ref{fig:f1}C, D. Similar to \citet{hah03} and \citet{tou02}, both $\psi_1$ and $\psi_2$ smoothly converge to their expected unsoftened values in the limit of $S \to 0$ for various PL disc models (i.e. $p$ and $q$). Further tests using other disc models, looking at the edge effects, etc. reinforce this conclusion.

This discussion strongly suggests that for any adopted form of softening, the expansion of the secular disturbing function must be performed following a certain rigorous procedure
\footnote{An analogous method is to modify the \textit{literal expansion} of disturbing function \citep[see][Ch. 6]{mur99} to account for softened interactions \citep[e.g. Tr98,][H03]{lee18}. This could be done by replacing $b_{1/2}^{(m)}$ with $\mathcal{B}_{1/2}^{(m)}$ in Eq. (7.1) of \citet{mur99} before applying the derivatives with respect to $\alpha$. We note that this procedure could apply for all $\mathcal{F}(r_1,r_2)$ with continuous first derivatives satisfying $D_1 + D_2 = -1$; see Appendix A.} 
as done, for instance, in Appendix \ref{app:plummer}. In other words, a direct replacement of the classical Laplace coefficients $b_{3/2}^{(m)}$ in Eq. (\ref{eq:deltaR}) with their softened analogues is, evidently, not sufficient for obtaining a well-behaved softened version of Laplace-Lagrange theory for co-planar discs.

\subsection{Implications for numerical applications} \label{sec:numerical_modeling}

In numerical studies of secular dynamics, self-gravitating discs are often treated as a collection of $N$ eccentric annuli (rings), with prescribed spacing (justified by the constancy of the semi-major axis), interacting gravitationally with each other \citep[e.g.][]{tou09, bat12}. This representation approximates a continuous particulate or fluid disc in the limit of $N \rightarrow \infty$. 

Computational cost associated with the evaluation of mutual ring-ring interactions in this setup, going as $\mathcal{O}(N^2)$, imposes limitations on the number of rings that can be used in practice. This is typically not a problem for the un-softened calculations, which converge to the expected full disc result even with a relatively coarse radial sampling of the integral contribution to e.g. the precession rate. Indeed, purple curves in Figures \ref{fig:f2} \& \ref{fig:f3} demonstrate this by showing the un-softened $\tilde\psi_1(x)$ and $\tilde\psi_2(x)$ computed without accounting\footnote{Note that, technically, in the un-softened case this mathematical procedure is not equivalent to introducing an actual physical gap in the disc, as the latter would result in additional boundary terms.} for the contributions from $a_{d,i} <a_p<a_{d,o}$ (see \S \ref{ref:spacing_cumulative}) to the integral terms in the un-softened expressions of \citet{irina18a}. These curves converge to the correct full disc result without exhibiting large variations in  $\tilde\psi_1(x)$ and $\tilde\psi_2(x)$, typical for softened cases. 

On the contrary, the results for the softened gravity presented in \S \ref{ref:spacing_cumulative} do elicit concern about the number of rings $N$ that is needed to accuratly capture the eccentricity dynamics of continuous razor-thin discs. Indeed, Figs. \ref{fig:f2} and \ref{fig:f3} reveal that the expected secular dynamics can be recovered using various softened gravity prescriptions only when one properly accounts for the gravitational effects of all disc annuli, including those very close to the orbit of particle under consideration. Indeed, we demonstrated that to reproduce both the magnitude and the sign of e.g. the precession rate, the distance $\Delta a$ separating a given test-particle orbit from nearest neighboring inner and outer disc rings should be quite small, $\Delta a /a_p \lesssim 0.1 \varsigma^2$. Only then does the delicate cancellation of large (in magnitude) contributions produced by different parts of the disc recovers the expected result. Thus, the separation between the modeled disc rings has to be substantially lower than the softening length itself ($\varsigma a_p$), meaning that $N$ has to be very large, $N\gtrsim 10\varsigma^{-2}$. This could easily make numerical studies of the eccentricity dynamics in discs very challenging.

\begin{figure}
	\includegraphics[width=1\columnwidth]{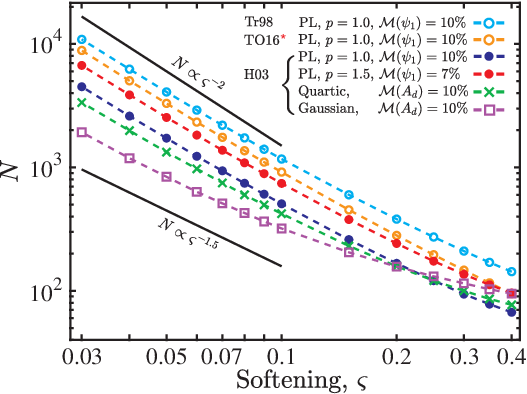}
   \caption{
   Scaling of number of softened annuli (rings) $N$ with softening parameter $\varsigma$ to ensure convergence of disc-driven free precession $A_d$ (or $\psi_1$) in discretized discs to the expected results in continuous \textit{softened} discs (Eqs. \ref{eq:An_general}, \ref{eq:psi1PL}). Calculations assume axisymmetric disc models extending from $a_{\rm in} = 0.1$ to $a_{\rm out} = 5$ AU: two PL discs (specified by $p$), a Quartic disc (same as Fig. \ref{fig:f5}) and a Gaussian ring (same as Fig. \ref{fig:f6}). We have used the softening methods of \citet{hah03}, \citet{tre98} and (corrected) \citet{teygor16}, as specified in the panel. Convergence is measured using the metric $\mathcal{M}(f)$ defined by Eq. (\ref{eq:metric}). One can see that, when $\varsigma \lesssim 0.1$, $N \sim C\varsigma^{-\beta}$, with $C\sim 10$ and $1.5\lesssim \chi\lesssim 2$. Similar results can be obtained for eccentric discs, and other softening prescriptions. See text (\S \ref{sec:numerical_modeling}) for details.
}
   \label{fig:f9}
\end{figure}

We further confirmed this expectation by studying the convergence of disc-driven free precession rate in numerically \textit{discretized softened} discs to the precession rate $A_d$ computed exactly for \textit{continuous softened} discs (Eqs. \ref{eq:An_general}, \ref{eq:psi1PL}).
To this end, we represented a given disc model as a collection of $N$ logarithmically-spaced rings, and measured the agreement between the radial profiles of theoretical and numerical results for $A_d$ (or $\psi_1$ for PL discs) by using the following global metric\footnote{For PL discs, we neglect rings within $10\%$ of disc edges when computing $\mathcal{M}(\psi_1)$.} 
\begin{equation}
    \mathcal{M}(f) = \sqrt{ \frac{\int_{a_{\rm in}}^{a_{\rm out}} [ f_{\rm theor}(a) - f_{\rm num}(a)    ]^2 da}{\int_{a_{\rm in}}^{a_{\rm out}}  f_{\rm theor}^2(a) da   }  }. 
    \label{eq:metric}
\end{equation}
Here $f_{\rm num}(a_i)$ is the value of the metric basis (e.g. precession rate $A_d$) evaluated at the position $a_i$ of $i$th ring by summing up the contributions of all other rings in the disc, while $f_{\rm theor}(a_i)$ is the analogous quantity computed in the limit of a continuous disc, i.e. as $N\to \infty$ (it is given by the non-discretized version of Eq. (\ref{eq:An_general}) if $f=A_d$, or Eq. (\ref{eq:psi1PL}) if $f=\psi_1$). Repeating this calculation for various combinations of $(N, \varsigma)$, we can determine the smallest number of rings $N(\varsigma)$ that ensures the desired convergence to within, e.g. $\sim 10\%$ (i.e. $\mathcal{M}(f) \sim 0.1$), for a given value of softening $\varsigma$. 

Figure \ref{fig:f9} depicts a sample of the results obtained using the softening methods of \citet{hah03}, \citet{tre98} and (rectified) \citet{teygor16} (see \S \ref{sect:gordon}) 
for various axisymmetric disc models as indicated in the legend\footnote{We exclude the softening method of \citet{tou02} from this analysis as it introduces additional complexity due to the nature of softening parameter; $\epsilon^2 = b^2 / \textrm{max}(a_1^2,a_2^2)$, see \S \ref{sect:touma}.}. Figure \ref{fig:f9} shows that as $\varsigma \to 0$, the number of rings scales as $N \sim C\varsigma^{-\chi}$ with\footnote{For example, the curve computed using the (corrected) model of \citet{teygor16} has $C = 10.9$ and $\chi = 1.91$, while the one for Quartic disc has $C=7.2$ and $\chi=1.75$.} $C\sim 10$ and $\chi\approx (1.8-1.9)$. The only notable exception is the Gaussian ring, for which convergence is faster (i.e. $N \propto \varsigma^{-1.5}$), probably because of mass concentration in a narrow range of radii. 

We note that the proportionality constant $C$ in the $N(\varsigma)$ relation is not perfectly defined in the sense that it depends on the (i) desired accuracy (roughly inversely proportional to $\mathcal{M}(f)$), (ii) adopted metric of accuracy (mild dependence), and (iii) softening prescription used --- Fig. \ref{fig:f9} shows that discretized calculations using softening model of \citet{hah03} require substantially lower (by $\sim 2$) number of annuli than those using the models of \citet{teygor16} and \citet{tre98}. Nevertheless, these results further reinforce the requirement of large number of rings, with $N \sim \varsigma^{-2}$, to capture the expected secular eccentricity dynamics in nearly-Keplerian discs.

Qualitatively similar results were stated in \citet{hah03} who showed that the secular effects of a continuous disc can be recovered only when the disc rings are sufficiently numerous that their radial separation is below the softening length. Although, interestingly, \citet{hah03} and \citet{lee18} claimed good convergence of the precession rate to the expected value already for $N\sim \mathcal{O}(\varsigma^{-1})$ (however, note that \citet{lee18} also included effects of gas pressure in their calculations, in addition to disc gravity). In our case, the condition on the separation between disc rings motivated by Figs. \ref{fig:f2} \& \ref{fig:f3} (i.e. $\Delta a/a_p \lesssim 0.1 \varsigma^2$), along with the results presented in Fig. \ref{fig:f9}, indicate that accurate representation of eccentricity dynamics in a cold, razor-thin disc requires a very large number of rings $N$ whenever small values of the softening parameter  are used.

As we have shown in \S \ref{sec:edgeeffects}, very small values of softening $\varsigma \lesssim 10^{-3}$ are, in fact, necessary to accurately capture eccentricity dynamics near the sharp edges of thin discs. This suggests that $N$ has to be prohibitively large when softened gravity is applied e.g. to study the dynamics of planetary ring \citep{GT79,chiang00,pan16}, which are known to have sharp edges. 


\subsection{Further generalizations and extensions}


All calculations in this work are based on the expansion of the secular disturbing function $R_d$ due to a coplanar disc --- softened and unsoftened --- to second order in eccentricities. This approximation may yield inaccurate results when the disc or particle eccentricities are high, e.g. in the vicinity of secular resonances where $A_d(a_p) = 0$ \citep{irina18a}, see Figs. \ref{fig:f5}, \ref{fig:f6}. Such situations may necessitate  a higher-order extension of the disc potential.

Such an exercise was pursued recently by \citet{sef18} who presented a calculation of  $R_d$ to $4$th order in eccentricities based on the un-softened method of \citet{hep80}. The general framework for calculating $R_d$ with arbitrary softening prescriptions presented in Appendix \ref{app:plummer} can also be extended to higher order in eccentricities in similar way\footnote{Another way to calculate the softened disturbing function for arbitrarily high eccentricities is to numerically compute  the ring-ring interaction potential, as was done by \citet{tou09}.}, see e.g. \citet{tou12}. We expect that conclusions similar to those drawn from our analysis in  \S \ref{sec:PL}-\ref{sec:edgeeffects} will also apply to the higher-order expansions. 

Additionally, although we only analyzed coplanar configurations in this work, the general framework presented in Appendix \ref{app:plummer} may be extended to account for non-coplanar configurations and study the inclination dynamics.

\section{Summary} 
\label{sec:summary}

In this work we investigated the applicability of softened gravity for computing the orbit-averaged potential of razor-thin eccentric discs. We compared disc-driven secular dynamics of coplanar test-particles computed using softening prescriptions available in the literature with the calculations based on the unsoftened method of \citet{hep80}. Our findings are summarized below.
\begin{itemize}
\item We confirmed that the softening methods of both  \citet{tou02} and \citet{hah03} correctly reproduce eccentricity dynamics of razor-thin discs in the limit of vanishing softening parameter $\varsigma$ for all disc models.
\item The softening prescription proposed in \citet{tre98} yields convergent results as $\varsigma \rightarrow 0$. However, quantitative differences of up to $\sim (20 - 30) \%$ from the unsoftened calculations are observed. We demonstrate that these differences arise because of the insufficient smoothness of the inter-particle interaction assumed in \citet{tre98}.
\item The softening formalism suggested in \citet{teygor16} does not result in convergent results in the limit of zero softening. 
\item Very small values of the (dimensionless) softening parameter are required for correctly reproducing secular eccentricity dynamics near sharp edges of disks/rings. 
\item We developed a general analytical framework for computing the secular disturbing function between two co-planar rings with arbitrary interaction potential of rather general form (Eq. \ref{eq:FS2}). This framework accurately reproduces the orbit-averaged razor-thin disc potential as $\varsigma \to 0$ for a wide class of softened gravity models. 
\item Using this general framework, we demonstrated that an accurate implementation of the softened potentials suggested in both \citet{tre98} and \citet{teygor16} leads to the recovery of the expected dynamical behavior in the limit of small softening. 
\item Our results suggest that the numerical treatments of the secular eccentricity dynamics in softened, nearly-Keplerian discs must obey important constraints. Namely, a fine numerical sampling (i.e. large number $N$ of discrete annuli representing the disc, with $N \sim C\varsigma^{-\chi}$, $C\sim O(10)$, $1.5\lesssim \chi \lesssim 2$) is required to ensure that the correct secular behavior is properly captured by such calculations when  $\varsigma$ is small. This finding has important ramifications for numerical treatments of planetary rings with sharp edges. 
\end{itemize}
In the future our results for the disc-driven eccentricity dynamics may be extended to higher order in eccentricity, as well as generalized for treating inclination dynamics.

\section*{Acknowledgements}

We express our gratitude to Scott Tremaine and Jihad Touma for a number of stimulating discussions, which have led to substantial improvements of the manuscript. We are also grateful to Gordon Ogilvie, Jean Teyssandier, Yoram Lithwick, and Cristobal Petrovich for useful discussions, and an anonymous referee for constructive comments. A.A.S. acknowledges a scholarship by the Gates Cambridge Trust (OPP1144), while R.R.R. was supported by NASA via grant 15-XRP15-2-0139. Open Access for this article was funded by the Bill \& Melinda Gates Foundation.


\bibliographystyle{mnras}
\bibliography{SR19_Softening_References}

\begin{landscape}
\begin{table}
	\caption{
	The coefficients $\phi_{ij}(\alpha)$ of the secular disturbing function with softened gravity featured in Eqs. (\ref{eq:An_general})-(\ref{eq:An_Bn_general}), which govern the individual secular ring-ring interaction (Eq. \ref{eq:R_softened_phi}), adopted from the literature (listed in the first column). Here $\alpha$ is defined such that $\alpha = a_< / a_>$ where $a_> = \text{max}(a_1,a_2)$ and $a_< = \text{min}(a_1,a_2)$. The softened interactions under consideration are those of \citet{tre98}, \citet{tou02}, \citet{hah03} and \citet{teygor16} -- see \S \ref{sec:softened_summary} for further details. For reference, the expressions of $\phi_{ij}^{\rm LL}$ corresponding to the (unsoftened) Newtonian ring-ring interaction (i.e. classical Laplace-Lagrange formalism) are also shown in the top row. The Laplace coefficients which are softened by the introduction of a softening parameter $\epsilon^2(\alpha)$ are defined in Eq. (\ref{eq:Bsm_softened}). Note that the expressions of $\phi_{ij}$ reported in \citet{tou02} have been corrected in a subsequent paper of \citet{tou12}. 
	}
	\label{table:table1}
	\begin{tabular}{lcccc} 
\hline
Formalism & $\epsilon^2(\alpha) $ & $\phi_{11}  $ & $\phi_{12} $ & $\phi_{22} $ 
\\
\hline
Laplace-Lagrange & --	&  $ \frac{1}{8}\alpha b_{3/2}^{(1)}  $ &$ -\frac{1}{4}\alpha b_{3/2}^{(2)}  $&  $\phi_{11}$ 
\\ \\
\citet{tre98} (Tr98)
& $\beta_c^2$
& $\frac{1}{8} \bigg(2\alpha \frac{d}{d\alpha} + \alpha^2 \frac{d^2}{d\alpha^2} \bigg) \mathcal{B}_{1/2}^{(0),\mathrm{Tr}}  $
&  $\frac{1}{4} \bigg(2 - 2\alpha \frac{d}{d\alpha} - \alpha^2 \frac{d^2}{d\alpha^2} \bigg) \mathcal{B}_{1/2}^{(1),\mathrm{Tr}}  $
& $\phi_{11}$ 
\\\\

&
& $ = \frac{1}{8}\alpha  \bigg[ \mathcal{B}_{3/2}^{(1),\mathrm{Tr}} - 3\alpha \beta_c^2 \mathcal{B}_{5/2}^{(0),\mathrm{Tr}}   \bigg]  $  
& $ =  -\frac{1}{4} \alpha \bigg[ \mathcal{B}_{3/2}^{(2),\mathrm{Tr}} - 3\alpha \beta_c^2 \mathcal{B}_{5/2}^{(1),\mathrm{Tr}}    \bigg] $ 
&
\\ \\
\citet{tou02} (T02)
& $ \beta^2 = b_c^2/ a_>^2 $ 
& $  -\frac{5}{8}\alpha \mathcal{B}_{3/2}^{(1),\mathrm{T}} +\frac{3}{16} \alpha^2 \mathcal{B}_{5/2}^{(0),\mathrm{T}} + \frac{3}{8}\alpha (1+\alpha^2) \mathcal{B}_{5/2}^{(1),\mathrm{T}}   $  
& $   \frac{9}{8}\alpha \mathcal{B}_{3/2}^{(0),\mathrm{T}}  + \frac{1}{8}\alpha \mathcal{B}_{3/2}^{(2),\mathrm{T}} - \frac{9}{8}\alpha(1+\alpha^2) \mathcal{B}_{5/2}^{(0),\mathrm{T}}          $  
& $  -\frac{5}{8}\alpha \mathcal{B}_{3/2}^{(1),\mathrm{T}} +\frac{3}{16} \alpha^2 \mathcal{B}_{5/2}^{(0),\mathrm{T}} + \frac{3}{8}\alpha (1+\alpha^2) \mathcal{B}_{5/2}^{(1),\mathrm{T}}   $ 
\\ \\
&
& $ -\frac{15}{16} \alpha^2 \mathcal{B}_{5/2}^{(2),\mathrm{T}} -\frac{3}{8}\alpha\beta^2 ( \alpha \mathcal{B}_{5/2}^{(0),\mathrm{T}} -  \mathcal{B}_{5/2}^{(1),\mathrm{T}}      ) $  
& $+ \frac{21}{16}\alpha^2 \mathcal{B}_{5/2}^{(1),\mathrm{T}}    + \frac{3}{8}\alpha (1+\alpha^2) \mathcal{B}_{5/2}^{(2),\mathrm{T}} + \frac{3}{16} \alpha^2\mathcal{B}_{5/2}^{(3),\mathrm{T}} $ 
&   $ -\frac{15}{16} \alpha^2 \mathcal{B}_{5/2}^{(2),\mathrm{T}} -\frac{3}{8} \beta^2 ( \mathcal{B}_{5/2}^{(0),\mathrm{T}} -  \alpha \mathcal{B}_{5/2}^{(1),\mathrm{T}})  $ 
\\ \\
\citet{hah03} (H03)
& $H^2 (1+\alpha^2) $
& $ \frac{1}{8} \alpha \bigg[  \mathcal{B}_{3/2}^{(1),\mathrm{H}} - 3\alpha H^2 (2+H^2) \mathcal{B}_{5/2}^{(0),\mathrm{H}}   \bigg]   $  
& $ -\frac{1}{4} \alpha \bigg[ \mathcal{B}_{3/2}^{(2),\mathrm{H}} - 3\alpha H^2 (2+H^2) \mathcal{B}_{5/2}^{(1),\mathrm{H}}   \bigg] $ 
& $\phi_{11}$ 
\\ \\
\citet{teygor16} (TO16)
& $S^2 \alpha$ 
& $ \frac{1}{8}\alpha\mathcal{B}_{3/2}^{(1),\mathrm{TO}}  $ 
&$ -\frac{1}{4}\alpha \mathcal{B}_{3/2}^{(2),\mathrm{TO}}  $
&  $\phi_{11}$ 
\\
\hline
	\end{tabular}
\end{table}
\end{landscape}


\appendix
\onecolumn

\section{Calculation of the secular ring-ring interaction} \label{app:plummer}

Here we present a calculation of the secular disturbing function due to two co-planar rings interacting with each other via softened gravity in the form (\ref{eq:FS2}). We do not assume any specific form for the softening function $\mathcal{F}$ apart from requiring it to be a function of the instantaneous positions of interacting particles with respect to the centre of the system. We first write the ring-ring interaction function as\footnote{Note that we do not deal with the indirect part of the potential -- which is left unsoftened -- as it contains only periodic terms and does not affect the secular dynamics \citep{mur99}.}
\begin{equation}
\Psi  = \bigg[ (\mathbf{r}_1 - \mathbf{r}_2)^2 + \mathcal{F}(r_1, r_2) \bigg]^{-1/2} 
= \bigg[ r_1^2 + r_2^2 - 2 r_1 r_2 \cos(f_1 - f_2 + \varpi_1 - \varpi_2) + \mathcal{F}(r_1, r_2) \bigg]^{-1/2},
 \label{eq:softenedinteraction}
\end{equation}
where $\mathcal{F}(r_1, r_2)$ is an arbitrary softening function introduced to cushion the singularity which arises otherwise at null inter-particle separations. In the above expression, $f_i$ is the true anomaly of the $i^{th}$ ring, $\varpi_i$ is its longitude of periapse and $r_i$ is its instantaneous position, $i=1,2$.
Our goal is to obtain the orbit-averaged expansion of $\Psi$ to second order in eccentricities $e_i$ valid for arbitrary $\mathcal{F}(r_1, r_2)$.


\subsection{Expansion of the interaction function $\Psi$ around small eccentricities}


Following the classical techniques of celestial mechanics \citep[see,][Ch. XVI]{plu18}, we start by expanding $\Psi$ around circular orbits. Using Taylor expansion we write
\begin{eqnarray}
\Psi &=&   \text{exp}\bigg\{ \log\bigg(\frac{r_1}{a_1} \bigg)D_1 + \log\bigg(\frac{r_2}{a_2} \bigg)D_2 + (f_1-M_1)D_3 + (f_2-M_2)D_4   \bigg\}  \Psi_0
\quad \equiv  \quad \mathbb{T} \Psi_0
\label{eq:Tdef}
\end{eqnarray}
with 
\begin{eqnarray}
\Psi_0 &=& \bigg[ a_1^2 + a_2^2 - 2 a_1 a_2 \cos\theta + \mathcal{F}(a_1, a_2) \bigg]^{-1/2},
\end{eqnarray}
where $\theta = M_1 - M_2 + \varpi_1 - \varpi_2$, $M_i$ represents the mean anomaly of the $i^{\rm th}$ ring characterized with semi-major axis $a_i$, and the linear operators $D_k$ are given by \citep{plu18}
\begin{equation}
D_1 = a_1 \frac{\partial}{\partial a_1} \equiv a_1 \partial_1  \quad, 
\qquad
D_2 = a_2 \frac{\partial}{\partial a_2}   \equiv a_2 \partial_2  \quad,
\qquad
\text{and} 
\qquad
D_3 = -D_4 = \frac{\partial}{\partial \theta}.
\end{equation}
Note that this expansion, as well as subsequent steps, is completely symmetric with respect to interchanging the particle indices.

Next, in order to calculate the action of the operator $\mathbb{T}$ defined by Eq. (\ref{eq:Tdef}) on the disturbing function of circular softened rings $\Psi_0$, we make use of the elliptical expansions of $r/a$ and $f-M$,
\begin{eqnarray}
(a^{-1} r )^D  &=&  1 - e  \cos M \cdot D + \frac{1}{2} e^2  [ 1-\cos(2M) ] \cdot D   + \frac{1}{4} e^2  [ 1+ \cos(2M) ] \cdot D (D-1)  + \mathcal{O}(e^3),
\\
\text{exp}\{(f-M) D \} &=& 1 + 2 e \sin M \cdot D + \frac{5}{4} e^2   \sin(2M) \cdot D  + e^2   [1-\cos(2M)] \cdot D^2  +  \mathcal{O}(e^3)
\end{eqnarray}
to multiply individual terms appearing in $\mathbb{T}$, keep the ones up to second order in eccentricities, and drop all terms which do not contain the difference of mean anomalies, $k(M_1-M_2)$, as they are evidently periodic and vanish upon orbit-averaging. Performing this procedure and dropping an irrelevant constant term, one can demonstrate that $\Psi$ reduces to 
\begin{eqnarray}
\Psi &=& \mathbb{T} \Psi_0 
\equiv \mathbb{A}\Psi_0 ~ e_1^2 + \mathbb{B}\Psi_0 ~ e_2^2 + \mathbb{C}\Psi_0 ~ e_1 e_2 \cos(\varpi_1-\varpi_2),
\label{eq:Psi}
\end{eqnarray}
where the operators $\mathbb{A}$, $\mathbb{B}$ and $\mathbb{C}$ acting on $\Psi_0$ are defined as
\begin{eqnarray}
\mathbb{A} &\equiv&  D_3^2 + \frac{1}{4} D_1(D_1+1),~~~~~~~~~
\mathbb{B} \equiv  D_4^2 + \frac{1}{4} D_2(D_2+1),
\label{eq:ABappendix}
\\
\mathbb{C} &\equiv&   \cos\theta \bigg( 2D_3 D_4 + \frac{1}{2} D_1 D_2 \bigg)   -\sin\theta (D_2 D_3 - D_1 D_4 ).   
\label{eq:Cappendix}
\end{eqnarray}
We have used the fact that $\cos(M_1-M_2) = \cos\theta \cos(\varpi_1- \varpi_2)$ and $\sin(M_1-M_2) = \sin\theta \cos(\varpi_1 - \varpi_2)$ in the secular regime \citep{plu18}.


\subsection{Computation of the action of relevant operators}


Equipped with the expression (\ref{eq:Psi}) for $\Psi$, we proceed to compute the action of operator $\mathbb{T}$ on $\Psi_0$ prior to orbit-averaging the resultant expression.
With this in mind, we compute the action of several operators appearing in the definitions of $\mathbb{A}$, $\mathbb{B}$ and $\mathbb{C}$ on $\Psi_0$ and list them below:
\begin{eqnarray}
D_3^2 ~ \Psi_0 &=&  D_4^2 ~ \Psi_0 \quad = \quad 3 a_1^2 a_2^2 \sin^2\theta ~ \Psi_0^{5} - a_1 a_2 \cos\theta ~ \Psi_0^{3},
\label{eq:D3squared}
\\
D_1 D_2 ~  \Psi_0 &=&  a_1 a_2 \bigg( \cos \theta - \frac{1}{2} \partial_1 \partial_2 \mathcal{F} \bigg)  ~ \Psi_0^{3}
+ 3  \bigg(a_2^2 - a_1 a_2 \cos\theta + \frac{a_2}{2} \partial_2 \mathcal{F} \bigg) \bigg( a_1^2 - a_1 a_2 \cos\theta + \frac{a_1}{2} \partial_1\mathcal{F}\bigg) ~ \Psi_0^{5},
\\
D_2 D_3 ~ \Psi_0 &=&  - a_1 a_2 \sin\theta ~ \Psi_0^{3} + 3  a_1 a_2 \sin\theta \bigg(a_2^2 - a_1 a_2 \cos\theta + \frac{a_2}{2} \partial_2 \mathcal{F} \bigg)~ \Psi_0^{5},
\\
D_1 D_4 ~ \Psi_0  &=&  a_1 a_2 \sin\theta ~ \Psi_0^{3} - 3  a_1 a_2 \sin\theta  \bigg(a_1^2 - a_1 a_2 \cos\theta + \frac{a_1}{2} \partial_1\mathcal{F} \bigg) ~\Psi_0^{5},
\\
D_1~ \Psi_0^3  &=& -3  \bigg( a_1^2 - a_1 a_2 \cos\theta + \frac{a_1}{2} \partial_1 \mathcal{F} \bigg) ~\Psi_0^5,
\\
D_2 ~ \Psi_0^3  &=& -3  \bigg( a_2^2 - a_1 a_2 \cos\theta + \frac{a_2}{2} \partial_2 \mathcal{F} \bigg) ~\Psi_0^5,
\end{eqnarray}
where for conciseness we have written $\mathcal{F}$ instead of $\mathcal{F}(a_1,a_2)$. 
Here, it is worthwhile to mention that, as far as the expansion technique is concerned, the terms $\partial_i \mathcal{F} ~(\text{with}~i=1,~2)$ appearing in the above expressions are the only difference brought upon by softening the Newtonian point-mass interaction (Eq. \ref{eq:softenedinteraction}).
Another set of operators useful in computing $\mathbb{T} \Psi_0$ is the following:
\begin{eqnarray}
D_1 (D_1 + 1) ~ \Psi_0  &=& - D_1 D_2 ~ \Psi_0
 + \frac{1}{2} D_1 \bigg[    \bigg(2\mathcal{F} - a_1 \partial_1\mathcal{F}  - a_2 \partial_2\mathcal{F} \bigg) ~ \Psi_0^3  \bigg], 
 \label{eq:D1D1}
\\
D_2 (D_2 + 1) ~ \Psi_0  &=& - D_1 D_2 ~ \Psi_0
 + \frac{1}{2}  D_2 \bigg[ \bigg( 2\mathcal{F} - a_1 \partial_1\mathcal{F}  - a_2 \partial_2\mathcal{F} \bigg) ~ \Psi_0^3 \bigg], 
 \label{eq:D2D2}
\end{eqnarray}
which can be obtained by making use of the identity 
$(D_1 + D_2 + 1 )\Psi_0 = \frac{1}{2}(2 \mathcal{F}- a_1 \partial_1 \mathcal{F} - a_2 \partial_2\mathcal{F})\Psi_0^3$. Here, we note that for all softening functions $\mathcal{F}$ for which 
$2\mathcal{F} - a_1 \partial_1\mathcal{F} - a_2 \partial_2 \mathcal{F} = 0$, one finds $D_1 + D_2 = -1$. Consequently, in such cases, the operators $D_1 (D_1+1)$ and $D_2(D_2+1)$ become identical rendering $\mathbb{A}\Psi_0 = \mathbb{B}\Psi_0$ (since $D_3^2 = D_4^2$, see Eqs. (\ref{eq:ABappendix}) and (\ref{eq:D3squared})). As a result, the resultant orbit-averaged disturbing function (\ref{eq:Psi}) is symmetric in $e_1$ and $e_2$, similar to the case of classical Laplace-Lagrange theory. This is not true in general, for instance, when $\mathcal{F}(r_1,r_2) =$ const $\neq 0$. 


\subsection{Orbit-averaging the interaction function $\Psi$}


The expressions (\ref{eq:D3squared})-(\ref{eq:D2D2}) allow the computation of $\Psi = \mathbb{T} \Psi_0$, which needs to be time-averaged in order to recover the secular disturbing function. We do not show the cumbersome collated expression for $\mathbb{T}\Psi_0$ and proceed to the final step of orbit-averaging, which will conclude our derivation. In short, our goal is to compute 
\begin{equation}
\langle \Psi \rangle = \langle \mathbb{T} \Psi_0 \rangle = \frac{1}{2\pi} \int\limits_0^{2\pi} \mathbb{T}\Psi_0 ~ d\theta,
\end{equation}
which essentially reduces to computing the individual terms $\langle \mathbb{A} \Psi_0\rangle$, $\langle \mathbb{B} \Psi_0\rangle$ and $\langle \mathbb{C} \Psi_0\rangle$. At the outset, it is important to note that each of the terms appearing in $\mathbb{T}\Psi_0$ (through $\mathbb{A}\Psi_0$, $\mathbb{B}\Psi_0$ and $\mathbb{C}\Psi_0$, or the operators they entail) are proportional to $\cos(m\theta) \Psi_0^{2s}$. By making use of $\alpha = {a_<}/{a_>}$, where $a_< = {\rm min}(a_1,a_2)$ and $a_> = {\rm max}(a_1,a_2)$, this combination can be reduced to
\begin{equation}
\cos(m\theta) \Psi_0^{2s} = a_>^{-2s} \cos(m\theta) \bigg[1+\alpha^2 - 2 \alpha \cos\theta + a_>^{-2} \mathcal{F}(a_1,a_2) \bigg]^{-s}.
\end{equation}
For that reason, calculation of the orbit-averaged $\Psi$ (by integrating over $d\theta$) yields integrals of the form
\begin{equation}
\mathcal{B}_s^{(m)}(\alpha) \equiv \frac{2}{\pi} \int\limits_0^{\pi} \cos(m\theta) \big[ 1+\alpha^2 - 2 \alpha \cos\theta + \epsilon^2(\alpha)          \big]^{-s}    d\theta , 
\label{eq:appBSM}
\end{equation}
which is the generalization of the classical Laplace coefficients $b_s^{(m)}$ (recovered when $\mathcal{F}(a_1,a_2) = 0$, see Eq. \ref{eq:bsm}) with the dimensionless softening parameter 
\begin{equation}
\epsilon^2(\alpha) \equiv a_>^{-2} \mathcal{F}(a_1,a_2),
\label{eq:epsAPP}
\end{equation} 
see Eq. (\ref{eq:Bsm_softened}). Employing this notation, we present the simplified expressions of $\langle\mathbb{A}\Psi_0\rangle$, $\langle\mathbb{B}\Psi_0\rangle$ and $\langle\mathbb{C}\Psi_0\rangle$ obtained as a result of orbit-averaging: 
\begin{eqnarray}
a_> ~ \langle \mathbb{A} \Psi_0 \rangle (\alpha) \equiv \phi_{11}(\alpha) &=& 
\frac{\alpha}{2} \bigg\{ -\frac{5}{4} \mathcal{B}_{3/2}^{(1)} + \frac{3}{8}\alpha \mathcal{B}_{5/2}^{(0)} 
+ \frac{3}{4}(1+\alpha^2) \mathcal{B}_{5/2}^{(1)} - \frac{15}{8} \alpha \mathcal{B}_{5/2}^{(2)} + \frac{3}{8} T_2 \mathcal{B}_{5/2}^{(1)} - \frac{3}{16} T_5 \mathcal{B}_{5/2}^{(0)} 
\nonumber 
\\
&+&   \frac{1}{8} \bigg( T_3 + \alpha^{-1} T_4 \bigg) \mathcal{B}_{3/2}^{(0)}
- \frac{3}{8} T_1 \bigg(\frac{a_1}{a_2} \mathcal{B}_{5/2}^{(0)} - \mathcal{B}_{5/2}^{(1)} + \frac{1}{2} T_7 \mathcal{B}_{5/2}^{(0)}\bigg)    
 \bigg\},
 \label{eq:Aaverage}
\\
a_> ~ \langle \mathbb{B} \Psi_0 \rangle (\alpha) \equiv \phi_{22}(\alpha)  &=& \frac{\alpha}{2} \bigg\{ -\frac{5}{4} \mathcal{B}_{3/2}^{(1)} + \frac{3}{8}\alpha \mathcal{B}_{5/2}^{(0)} 
+ \frac{3}{4}(1+\alpha^2) \mathcal{B}_{5/2}^{(1)} - \frac{15}{8} \alpha \mathcal{B}_{5/2}^{(2)} + \frac{3}{8} T_2 \mathcal{B}_{5/2}^{(1)} - \frac{3}{16} T_5 \mathcal{B}_{5/2}^{(0)}
\nonumber 
\\
&+&  \frac{1}{8} \bigg( T_3 + \alpha^{-1} T_6 \bigg) \mathcal{B}_{3/2}^{(0)} 
- \frac{3}{8} T_1 \bigg(\frac{a_2}{a_1} \mathcal{B}_{5/2}^{(0)} - \mathcal{B}_{5/2}^{(1)} + \frac{1}{2} T_8 \mathcal{B}_{5/2}^{(0)}\bigg)    
 \bigg\},
  \label{eq:Baverage}
\\
a_> ~ \langle \mathbb{C} \Psi_0 \rangle (\alpha) \equiv \phi_{12}(\alpha)  &=& \frac{\alpha}{2} \bigg\{
\frac{9}{4} \mathcal{B}_{3/2}^{(0)} + \frac{1}{4} \mathcal{B}_{3/2}^{(2)} 
+ \frac{3}{8} \alpha \mathcal{B}_{5/2}^{(3)} + \frac{21}{8} \alpha \mathcal{B}_{5/2}^{(1)} + \frac{3}{4}(1+\alpha^2) \mathcal{B}_{5/2}^{(2)} - \frac{9}{4}(1+\alpha^2)\mathcal{B}_{5/2}^{(0)} 
\nonumber 
\\
&-& \frac{1}{4} T_3 \mathcal{B}_{3/2}^{(1)}  - \frac{9}{8} T_2 \mathcal{B}_{5/2}^{(0)} + \frac{3}{8} T_5 \mathcal{B}_{5/2}^{(1)} 
+ \frac{3}{8} T_2 \mathcal{B}_{5/2}^{(2)} 
\bigg\}.
 \label{eq:Caverage}
\end{eqnarray}
In equations  (\ref{eq:Aaverage})-(\ref{eq:Caverage}), we have defined the dimensionless functions $T_i(\alpha)$ such that
\begin{eqnarray}
T_1 &=& a_>^{-2} \big( 2 \mathcal{F} - a_1 \partial_1 \mathcal{F} - a_2 \partial_2 \mathcal{F} \big),
~~~~~~T_2 = \alpha \bigg(  \frac{\partial_1 \mathcal{F}}{a_2} + \frac{\partial_2 \mathcal{F}}{a_1} \bigg),
~~~~~~T_3 = \partial_1 \partial_2 \mathcal{F},
\label{eq:T123}
\\
T_4 &=& \frac{a_1}{a_>^{2}} \partial_1[2 \mathcal{F} - a_1 \partial_1 \mathcal{F} - a_2 \partial_2 \mathcal{F}],
~~~~~~~T_5 = \alpha \bigg( 2 \frac{\partial_1 \mathcal{F}}{a_1} + 2 \frac{\partial_2 \mathcal{F}}{a_2} + \frac{\partial_1 \mathcal{F}}{a_1} \frac{\partial_2 \mathcal{F}}{a_2} \bigg),
\\
T_6 &=& \frac{a_2}{a_>^{2}} \partial_2[ 2 \mathcal{F} - a_1 \partial_1 \mathcal{F} - a_2 \partial_2 \mathcal{F}],
~~~~~~~T_7 =  a_2^{-1} \partial_1 \mathcal{F},
~~~~~~~T_8 =  a_1^{-1} \partial_2 \mathcal{F},
\label{eq:T678}
\end{eqnarray}
where, as before, $\mathcal{F} \equiv \mathcal{F}(a_1,a_2)$, $\alpha = a_</a_>$ and $\partial_i \equiv \partial/\partial a_i$. Note that the expressions for $\phi_{11}$ and $\phi_{22}$ swap  definitions upon replacing $a_1$ by $a_2$, whilst keeping $\alpha < 1$ by construction. This can be understood by first noting that functions $T_i$ with $i= 1,~2,~3,$ and $5$ are invariant under $a_1 \leftrightharpoons a_2$ while, at the same time, $T_4$ and $T_7$ (appearing in the second line of Eq. (\ref{eq:Aaverage})) translate to $T_6$ and $T_8$ (appearing in the second line of Eq. (\ref{eq:Baverage})); and vice versa.

These identities, when combined,  yield the desired expression of $\langle \Psi \rangle = \langle \mathbb{T} \Psi_0 \rangle$; see Eqs. (\ref{eq:Psi})-(\ref{eq:Cappendix}). Subsequently, the softened ring-ring disturbing function in the form  (\ref{eq:R_softened_phi}) is recovered, with the coefficients $\phi_{ij}$ defined by Eqs. (\ref{eq:Aaverage}) -- (\ref{eq:Caverage}). This completes our calculation of the secular ring-ring interaction between two softened coplanar rings, up to second order in eccentricity and valid for arbitrary softening functions $\mathcal{F}(r_1,r_2)$. 

Note that in the absence of softening (i.e. $\mathcal{F}(r_1,r_2) = 0$) $T_i=0$ for all $i$ and the classical expressions for $\phi_{11}^{\rm LL}$,  $\phi_{22}^{\rm LL}$ and $\phi_{12}^{\rm LL}$ --- Eqs. (\ref{eq:classical_psi1})-(\ref{eq:classical_psi}) --- are recovered. Finally, we mention that the expansion technique exploited here can be used to recover the orbit-averaged disturbing function valid to arbitrary order in eccentricity, as well as inclinations.


\subsection{Alternative calculation: secular disc-particle interaction}
\label{app:disc}


Calculations presented above describe the orbit-averaged coupling between the two individual annuli, which subsequently need to be integrated over the semi-major axes of the disc elements to represent the effect of a continuous disc. In principle, one can also arrive at the expressions (\ref{eq:Rd_general}) by assuming a continuous mass distribution in the disc from the start and performing a calculation similar to that in \citet{irina18a}. Namely, one would need to compute $R_d= \langle G\int_S \Sigma({\bf r}_d)\Phi({\bf r}_d,{\bf r}_p) dS \rangle$, where $\Phi$ is the interaction potential given by equation (\ref{eq:FS2}), angle brackets indicate averaging over the orbit of the test particle given by ${\bf r}_p$ and integration is carried out over the full surface of the disc $S$ with ${\bf r}_d$ denoting the location of a disc element. To obtain the expression for $R_d$ accurate to second order in eccentricities one would need to expand $\Phi({\bf r}_d,{\bf r}_p)$ to second order in particle and disc eccentricities by e.g. writing $r_p=a_p(1-e_p\cos E_p)$, where $E_p$ is the eccentric anomaly of the particle orbit. This expansion should explicitly account for the dependence of $\mathcal{F}$ on ${\bf r}_d$ and ${\bf r}_p$. Averaging the resulting expressions over $E_p$, one would arrive at the proper expression for $R_d$ in the form (\ref{eq:Rd_general}).

In particular, after a lengthy but straightforward calculation this method gives the following expression for the disc-driven precession rate:
\ba
\mathsf{A}_d=\frac{\pi G}{2n_p a_p^2} \int\frac{a\Sigma(a)da}{a_>} &\Bigg\{ & 
\frac{1}{4}\left[3 a_p\mathcal{F}^\prime \left(\mathcal{F}^\prime+4a_p\right) - 2\left(2\mathcal{F}^\prime+a_p\mathcal{F}^{\prime\prime}\right)\left(a_p^2+a^2+\mathcal{F}\right)-12a_p\mathcal{F} \right]\frac{a_p\mathcal{B}_{5/2}^{(0)}(\alpha)}{a_>^4} 
\nonumber\\
&+& \alpha \mathcal{B}_{3/2}^{(1)}(\alpha)-\left(\mathcal{F}^\prime-a_p\mathcal{F}^{\prime\prime}\right)\frac{a_p \alpha \mathcal{B}_{5/2}^{(1)}(\alpha)}{a_>^2}\Bigg\},
\label{eq:disc_Ad}
\ea  
where prime denotes differentiation with respect to $a_p$ (e.g. $\mathcal{F}^\prime=\partial\mathcal{F}/\partial a_p$), $a_>=\max(a_p,a)$, $\alpha=\min(a_p,a)/\max(a_p,a)$ and integration is done over the semi-major axis $a$ of the disk elements. Calculation of the non-axisymmetric part of $R_d$ resulting from non-zero disk eccentricity (i.e. $\bm{\mathsf{B}}_d$) is somewhat more tedious but can nevertheless be done similar to \citet{irina18a}.


\section{ Specific Cases of $\mathcal{F}(r_1,r_2)$} 
\label{app:correctgordon}


\begin{table}
\centering
\caption{
	The functional forms of the coefficients $T_i(\alpha)$ given by Eqs. (\ref{eq:T123})-(\ref{eq:T678}) appearing in the orbit-averaged  disturbing function due to two coplanar (arbitrarily) softened rings (Eq. \ref{eq:Aaverage}-\ref{eq:Caverage}) such that $\alpha \equiv a_< / a_> \leq 1$. The first column lists the softening prescriptions analyzed in this work (see \S \ref{sec:softened_summary}), while the second column shows the specific forms of the softening function $\mathcal{F}(r_1,r_2)$ in Eq. (\ref{eq:softenedinteraction}). The corresponding expressions for the dimensionless softening parameters $\epsilon^2(\alpha) = a_>^{-2} \mathcal{F}(a_1,a_2)$ (Eq. \ref{eq:epsAPP}) entering in the definition of softened Laplace coefficients (Eq. \ref{eq:appBSM}) are also shown. Here, $\Theta(x)$ represents the Heaviside step function and $\delta(x) = d\Theta(x)/dx$ stands for Dirac delta-function.
}
	\label{table:tableApp}
	\begin{tabular}{lcclcclclcc} 
\hline
Method
   & $\mathcal{F}(r_1,r_2) $  & $\epsilon^2(\alpha)$ &$T_1(\alpha)$ & $T_2(\alpha)$ & $T_3(\alpha)$ & $T_4(\alpha)$ & $T_5(\alpha)$ & $T_6(\alpha)$ & $T_7(\alpha)$ & $T_8(\alpha)$
\\
\hline
H03
& $H^2(r_1^2 + r_2^2 )$ 
& $H^2 (1+\alpha^2)$
& $ 0 $ 
& $2 H^2 (1+\alpha^2)$
& $0$
& $0$
& $4 \alpha H^2 (2+H^2)$
& $0$
& $2 H^2 \frac{a_1}{a_2}$ 
& $2  H^2 \frac{a_2}{a_1}$  
\\ 
\\
T02
& $b_c^2$
& $\beta^2 = (b_c/a_>)^2$
& $ 2 \beta^2 $ 
& $0$
& $0$
& $0$
& $0$
& $0$
& $0$
& $0$
\\ 
\\
Tr98
& $\beta_c^2 \rm{max}(r_1^2,r_2^2)$
& $\beta_c^2$
& $ 0 $ 
& $2\beta_c^2$
& $-2 \beta_c^2 \delta(\alpha - 1)$
& $0$
& $4\alpha\beta_c^2$
& $0$
& $  \frac{2 \beta_c^2}{\alpha}  \Theta(a_1-a_2)$ 
& $  \frac{2 \beta_c^2}{\alpha}  \Theta(a_2-a_1)$ 
\\ 
\\
TO16
& $S^2 r_1 r_2 $
& $S^2\alpha$
& $ 0 $ 
& $2\alpha S^2$
& $S^2$
& $0$
& $ S^2 ( S^2\alpha + 2\alpha^2 +2 )$
& $0$
& $S^2$
& $ S^2$
 \\
\hline
	\end{tabular}
\end{table}

The general framework developed in Appendix \ref{app:plummer} allows us to recover the expressions of $\phi_{ij}$ arrived at by \citet{tou02}  and \citet{hah03} upon specifying certain functional forms of $\mathcal{F}(r_1,r_2)$. 
Indeed, \citet{tou02} performed the same calculations as presented in Appendix \ref{app:plummer} for the case of Plummer potential -- $\mathcal{F}(r_1,r_2) = b_c^2$ --  to second order in eccentricities, and later to fourth order in eccentricities \citep{tou12}. 
Furthermore, we find that the results obtained by  \citet{hah03} can be recovered from our general framework by setting 
$\mathcal{F}(r_1,r_2) = H^2 (r_1^2 + r_2^2)$. For reference, the functional forms of $T_i$ for these forms of $\mathcal{F}(r_1,r_2)$, along with their softening parameters $\epsilon^2(\alpha)$, are summarized in Table \ref{table:tableApp}, which can be used to show that Eqs. (\ref{eq:Aaverage})-(\ref{eq:Caverage}) reduce to those in Table \ref{table:table1} after some algebra with the aid of the recursive relationships for $\mathcal{B}_s^{(m)}$ presented in Appendix \ref{app:lap}. 

As to the formalism of \citet{teygor16}, we find, using their softening prescription of $\mathcal{F}(r_1,r_2) = S^2 r_1 r_2$, that our general framework yields $\phi_{ij}$ expressions different from those reported by \citet{teygor16}. Indeed, we first note that in this case,  $T_1 = T_4 = T_6  = 0$ (Table \ref{table:tableApp}) rendering the expressions of $\phi_{11}$ and  $\phi_{22}$ identical such that 
\begin{equation}
\phi_{11} = \phi_{22} = \frac{\alpha}{8} \bigg\{ -5 \mathcal{B}_{3/2}^{(1),\mathrm{TO}} + \frac{3}{2}\alpha \mathcal{B}_{5/2}^{(0),\mathrm{TO}} 
+ 3 (1+\alpha^2 +S^2\alpha) \mathcal{B}_{5/2}^{(1),\mathrm{TO}} - \frac{15}{2} \alpha \mathcal{B}_{5/2}^{(2),\mathrm{TO}} - \frac{3}{4} S^2 (S^2\alpha + 2\alpha^2 +2) \mathcal{B}_{5/2}^{(0),\mathrm{TO}} + \frac{1}{2} S^2 \mathcal{B}_{3/2}^{(0),\mathrm{TO}}    \bigg\}
\label{eq:psi11psi22_mid}
\end{equation}
Using the recursive relationships listed in Appendix \ref{recursive}, the above expression can be simplified further. Indeed,  Eq. (\ref{eq:lap2}) with $m=1$ and $s=5/2$ and Eq. (\ref{eq:lap1}) with $m=1$ and $s=3/2$ read 
\begin{eqnarray}
3 (1+\alpha^2 +S^2\alpha) \mathcal{B}_{5/2}^{(1),\mathrm{TO}} &=& -\frac{3\alpha}{2} \mathcal{B}_{5/2}^{(2),\mathrm{TO}} + \frac{15}{2} \alpha \mathcal{B}_{5/2}^{(0),\mathrm{TO}},
\\
-6 \mathcal{B}_{3/2}^{(1),\mathrm{TO}} &=& 9 \alpha \bigg( \mathcal{B}_{5/2}^{(2),\mathrm{TO}} - \mathcal{B}_{5/2}^{(0),\mathrm{TO}}         \bigg),
\end{eqnarray}
respectively. Inserting the above two identities in Eq. (\ref{eq:psi11psi22_mid}) one arrives at Eq. (\ref{eq:new22}).
Similarly, the expression of $\phi_{12}$  (Eq. \ref{eq:Caverage}) can be simplified with the aid of Eq. (\ref{eq:lap3}) (with $m=0,~s=3/2$), Eq. (\ref{eq:lap2}) (with $m = 2,~ s= 5/2$) and Eq. (\ref{eq:lap1}) (with $m = 2,~ s= 3/2$) resulting in Eq. (\ref{eq:new2}) after some algebra. As discussed in \S \ref{sect:gordon}, the terms in Eqs. (\ref{eq:new22})-(\ref{eq:new2}) explicitly proportional to $S^2$ are absent in the original formulation of \citet{teygor16} (see Table \ref{table:table1}). 

Similarly, for the formalism of \citet{tre98}, propagating their functional form of $\mathcal{F}(r_1,r_2) = \beta_c^2 {\rm max}(r_1^2,r_2^2)$ through our general framework, we arrive at the expressions for $\phi_{ij}(\alpha)$ differing from those reported in \citet{tre98} in a very special way: we find $\phi_{ij}$ to contain additionl terms proportional to $T_3(\alpha) \sim \delta(\alpha-1)$, where $\delta(x)$ is the Dirac delta-function. Such terms are absent in the original formulation of \citet{tre98} (see Tables \ref{table:table1}, \ref{table:tableApp}). Emergence of these terms can be easily demonstrated by first noting that in this case $\phi_{11} = \phi_{22}$ (as $T_1 = T_4 = T_6 = 0$),  employing the recursive relationships for Laplace coefficients (in a similar order as done above for TO16) to simplify the general expressions of $\phi_{11} (= \phi_{22})$ and $\phi_{12}$, and finally arriving at Eqs. (\ref{eq:new_Tr_axi}), (\ref{eq:new_Tr_ecc}). The ramifications of this finding is discussed in Section \ref{sec:tremaine_why}.

\section{Generalized Laplace coefficients} 
\label{app:lap}

As demonstrated in Appendix \ref{app:plummer}, softening the Newtonian point-mass potential by an arbitrary function $\mathcal{F}(r_1,r_2)$ modifies the definition of the Laplace coefficients as shown by Eqs. (\ref{eq:Bsm_softened}), (\ref{eq:appBSM}) by the introduction of a softening parameter
$\epsilon^2(\alpha) = a_{>}^{-2} \mathcal{F}(a_1,a_2)$ 
(Eq. \ref{eq:epsAPP}), $0 \le \alpha = a_< /a_> \le 1$.
Here, we present some useful recursive relationships amongst different generalized Laplace coefficients $\mathcal{B}_s^{(m)}(\alpha)$, along with their asymptotic behavior in the limits of $\alpha \rightarrow 0,~ 1$ as well as their relationship to complete elliptic integrals.

\subsection{Recursive Relations} 
\label{recursive}

Generalizing the results for the usual (unsoftened) Laplace coefficients $b_{s}^{(m)}$ \citep[e.g.][p. 159]{plu18}, the following relationships can be easily obtained for the generalized Laplace coefficients defined by Eq. (\ref{eq:Bsm_softened}),(\ref{eq:appBSM}):
\begin{eqnarray}
m \mathcal{B}_{s}^{(m)}  &=& s\alpha   \mathcal{B}_{s+1}^{(m-1)} - s\alpha   \mathcal{B}_{s+1}^{(m+1)} ,       
\label{eq:lap1}
\\
m(1+\alpha^2 +\epsilon^2)  \mathcal{B}_{s}^{(m)} &=& \alpha (m+1-s) \mathcal{B}_{s}^{(m+1)} 
+ \alpha (m+s-1) \mathcal{B}_{s}^{(m-1)},
\label{eq:lap2}
\\
(m+s) \mathcal{B}_{s}^{(m)} &=& s(1+\alpha^2 +\epsilon^2) \mathcal{B}_{s+1}^{(m)} 
- 2s\alpha \mathcal{B}_{s+1}^{(m+1)}.
\label{eq:lap3}
\end{eqnarray}
The difference with the classical recursive relations for $b_{s}^{(m)}$ amounts to  substituting the combination $1+\alpha^2$ appearing in the case of ordinary Laplace coefficients with $1+\alpha^2 + \epsilon^2(\alpha)$.

Another useful expression relating the generalized Laplace coefficients of arguments $\alpha$ and $\alpha^{-1}$ is
\begin{equation}
\mathcal{B}_{s}^{(m)}(\alpha^{-1}) = \alpha^{2s} \mathcal{B}_{s}^{(m)}(\alpha).
\label{eq:Bsm_inversion}
\end{equation}
Note that the above relationship is valid only as long as  the softening parameter satisfies $\alpha^2 \epsilon^2(1/\alpha) = \epsilon^2(\alpha) $. For instance, this condition is violated when the softening parameter $\epsilon$ has no dependence on $\alpha$ (e.g. that of \citet{tre98}, see Table \ref{table:table1}).

\subsection{Asymptotic Behavior}

Here we derive approximate expressions for $\mathcal{B}_{s}^{(m)}$ in the asymptotic limits; for $\alpha \rightarrow 0$ and $\alpha \rightarrow 1$.
\\
{\bf Case 1:} In the limit of $\alpha \approx 0$, one can factor out the term $1+\alpha^2 + \epsilon^2(\alpha)$ from the integrand of $\mathcal{B}_{s}^{(m)}$ to expand the denominator around $\gamma^{-1} \approx 0$, where $\gamma=(2\alpha)^{-1}[1+\alpha^2 + \epsilon^2(\alpha)]$. This allows us to approximate $\mathcal{B}_{s}^{(m)}$ as 
\begin{eqnarray}
& B_{s}^{(m)}(\alpha) & 
 \approx  \frac{2}{\pi(2\alpha\gamma)^s} \int\limits_{0}^{\pi} \cos(m\theta) \times \bigg[1 + \frac{s}{\gamma}\cos\theta 
+ \frac{s(s+1)}{2\gamma^2} \cos^2\theta  + \frac{s(s+1)(s+2)}{6\gamma^3} \cos^3\theta   \bigg] d\theta.
\end{eqnarray}
Using the orthogonality of the cosine functions, it is straightforward to show that
\begin{equation}
\mathcal{B}_{s}^{(m)} \approx \frac{\alpha^m F_m}{(2\alpha\gamma)^{s+m}}, ~~~~~ \text{as ~~}\alpha \to 0, \text{~~where} ~~~F_m=\left\{
  \begin{array}{@{}ll@{}}
    2 & \text{if}\ m=0 \\
    2s & \text{if}\ m=1 \\
    s(s+1) & \text{if}\ m=2 \\
    \frac{1}{3}s(s+1)(s+2) & \text{if}\ m=3 \\
  \end{array}\right.
  \label{eq:mathcalB_alpha0}
\end{equation}

{\bf Case 2:} In the opposite limit of $x = 1-\alpha \approx 0$, the dominant contribution to $\mathcal{B}_{s}^{(m)}$ comes from $\theta\ll 1$ \citep{gol80}. Thus one can set $\cos(m\theta) \to 1$ in the numerator, approximate $\cos\theta \approx 1- \theta^2/2$ in the denominator and extend the integration limit to infinity. Furthermore, setting $\alpha = 1$ (i.e. $x = 0$) everywhere except when it appears in the combination $1-\alpha$, the generalized Laplace coefficient can be approximated as
\begin{equation}
\mathcal{B}_{s}^{(m)} \approx \frac{2}{\pi} \int\limits_{0}^{\infty} \frac{  d\theta}{\bigg[x^2 + \theta^2 + \epsilon^2_{\alpha=1} \bigg]^{s}} 
= \frac{2}{\pi} \left\{
  \begin{array}{@{}ll@{}}
    (x^2+  \epsilon^2_{\alpha=1})^{-1}   & \text{if}\ s = 3/2 \\
    (2/3) (x^2+\epsilon^2_{\alpha=1}  )^{-2}  & \text{if}\ s = 5/2 \\
  \end{array}\right.
    \label{eq:mathcalB_alpha1}
\end{equation}
where $\epsilon^2_{\alpha=1}$ is the softening parameter evaluated at $\alpha =  1$.

\subsection{Relationship to elliptic integrals} \label{app:ellipticintegrals}

Here we express the generalized Laplace coefficients $\mathcal{B}_s^{(m)}$ in terms of complete elliptic integrals. These expressions can be used for rapid numerical evaluation of the generalized Laplace coefficients without relying on numerical integration of Eq. (\ref{eq:appBSM}) (or Eq. (\ref{eq:Bsm_softened})). Let us write, as before, $2\alpha \gamma = 1+ \alpha^2 + \epsilon^2(\alpha)$ and define $\chi = \sqrt{2/(\gamma+1)}$ such that, for any general softening parameter $\epsilon^2(\alpha)$, we have $0\leq \chi \leq 1$ and $\gamma \geq 1$. Now let us express $\mathcal{B}_s^{(m)}$ in terms of $\gamma$ to write 
\begin{equation}
\mathcal{B}_s^{(m)} = \frac{2^{1-s}}{\pi \alpha^s}  \int\limits_{0}^{\pi} \frac{\cos(m\theta) } {(\gamma-\cos\theta)^{s}} d\theta.
\end{equation}
Introducing complete elliptic integrals
$\mathbf{K}(\chi) = \int_0^{\pi/2}  \left(1-\chi^2\sin^2\phi\right)^{-1/2}  d\phi$ 
~~and ~~
$\mathbf{E}(\chi) = \int_0^{\pi/2}  \left(1-\chi^2\sin^2\phi\right)^{1/2}  d\phi$, we find that 
\begin{eqnarray}
\mathcal{B}_{3/2}^{(0)} &=&  \frac{2 \mathbf{E}(\chi) }{ \pi \alpha (\gamma-1) \sqrt{2 \alpha (\gamma+1) } },   
\qquad \qquad \qquad \qquad ~~~~~~~~
\mathcal{B}_{3/2}^{(1)} = \frac{2 \bigg[ -(\gamma-1)\mathbf{K}(\chi)  +\gamma \mathbf{E}(\chi) \bigg]   }{ \pi \alpha (\gamma-1) \sqrt{2 \alpha (\gamma+1) } }   ,
\\
\mathcal{B}_{3/2}^{(2)} &=& \frac{2 \bigg[ -4\gamma(\gamma-1)\mathbf{K}(\chi)  +(4\gamma^2-3) \mathbf{E}(\chi) \bigg]   }{ \pi \alpha (\gamma-1) \sqrt{2 \alpha (\gamma+1) } }   ,
~~~~~~~~~~~~~
\mathcal{B}_{3/2}^{(3)} = \frac{2}{3} \frac{\bigg[ -(\gamma-1)(32\gamma^2-5)\mathbf{K}(\chi)  + \gamma (32\gamma^2-29) \mathbf{E}(\chi) \bigg]   }{ \pi \alpha (\gamma-1) \sqrt{2 \alpha (\gamma+1) } } ,
\\
\mathcal{B}_{5/2}^{(0)} &=& \frac{4 \bigg[ -(\gamma-1)\mathbf{K}(\chi)  +4\gamma \mathbf{E}(\chi) \bigg]   }{ 3\pi (2\alpha)^{5/2} (\gamma+1)^{3/2} (\gamma-1)^2   }   ,
~~~~~~~~~~~~~~~~~~~~~~~~~~~~~
\mathcal{B}_{5/2}^{(1)} = \frac{4 \bigg[ -\gamma (\gamma-1)\mathbf{K}(\chi)  +(\gamma^2+3) \mathbf{E}(\chi) \bigg]   }{ 3\pi (2\alpha)^{5/2} (\gamma+1)^{3/2} (\gamma-1)^2   }   ,
\\
\mathcal{B}_{5/2}^{(2)} &=& \frac{4 \bigg[ (\gamma-1)(4\gamma^2-5)\mathbf{K}(\chi)  -4\gamma(\gamma^2-2) \mathbf{E}(\chi) \bigg]   }{ 3\pi (2\alpha)^{5/2} (\gamma+1)^{3/2} (\gamma-1)^2   }   ,
~~~~~
\mathcal{B}_{5/2}^{(3)} = \frac{4 \bigg[ \gamma(\gamma-1)(32\gamma^2-33)\mathbf{K}(\chi)  
- (32\gamma^4-57 \gamma^2 +21) \mathbf{E}(\chi) \bigg]   }{ 3\pi (2\alpha)^{5/2} (\gamma+1)^{3/2} (\gamma-1)^2   }.   
\end{eqnarray}
These expressions permit efficient numerical evaluation of arbitrarily softened Laplace coefficients as functions of $\alpha$, since effective algorithms for computing $\mathbf{K}$ and $\mathbf{E}$ exist \citep{NR}.

\section{Convergence Criterion for the pre-factors of power-law discs} 
\label{app:convergence}

Astrophysical discs often extend over a few orders of magnitude in radius so that $a_{\rm out} / a_{\rm in} \gg 1$. In such situations, far from the disc edges one can take the limit of both $\alpha_1 =  a_{\rm in} /a_p$ and $\alpha_2 = a_p / a_{\rm out}$ going to zero, provided that the gravitational potential of a power-law disc is insensitive to the locations of the disc boundaries (see Eqs. \ref{eq:psi1PL}, \ref{eq:phi1phi2PL}). Then the pre-factors $\psi_1$ and $\psi_2$ of the disturbing function converge to values depending only on the power-law indices $p$ and $p+q$ respectively, as well as on the adopted softening prescription.

The conditions on the values of $p$ and $q$ which guarantee this convergence can be determined by expanding the coefficients $\phi_{ij}(\alpha)$, which appear in the integrands of each of $\psi_1$ and $\psi_2$, in the limit of $\alpha \approx 0$. Using the Taylor expansions of softened Laplace coefficients $\mathcal{B}_s^{(m)}$, we determined that both $\psi_1$ and $\psi_2$ calculated using the softening methods of \citet{hah03} and \citet{tre98}
(as well as its rectified version) are convergent as long as $ -1 < p < 4$ and $  -2 < p + q <5$, respectively, for all values of softening (i.e. $H, \beta_c$). This follows from the fact that for both \citet{hah03} and \citet{tre98} we have $\phi_{11} = \phi_{22}  \sim \alpha^2$ and $\phi_{12} \sim \alpha^3$ to lowest order in $\alpha$. These ranges of $p$ and $p+q$ are in line with the findings of \citet{sil15}.

As to the (rectified) softening model of \citet{teygor16}, a similar exercise yields that $\phi_{11} = \phi_{22} \approx -\frac{1}{4}S^2 \alpha +  \frac{3}{8} ( 1 + \frac{3}{2} S^4)   \alpha^2 $ and $\phi_{12} \approx \frac{3}{2}S^2 \alpha^2  - \frac{15}{16} ( 1+ 5 S^4) \alpha^3 $ which, in the limit of $S \to 0$, translate to the same ranges for $\psi_1$ and $\psi_2$ convergence as \citet{sil15}. However, when $S$ is relatively large, it is trivial to show that $\psi_1$ and $\psi_2$ are convergent over limited ranges of $0<p<3$ and $-1 < p+q < 4$, respectively. A similar analysis for the softening method of \citet{tou02} reveals that the ranges for $\psi_1$ and $\psi_2$ convergence are in line with the findings of \citet{sil15} when the corresponding softening parameter $b_c \to 0$. However, when $b_c$ is non-zero, the ranges are narrowed down to $-1< p < 2$ and $-2 < p+q < 3$ respectively.

\bsp	
\label{lastpage}
\end{document}